\documentclass[12pt,aps,nofootinbib]{revtex4}
\usepackage[utf8]{inputenc}
\usepackage{amsmath}
\usepackage{amsfonts}
\usepackage{amssymb}
\usepackage{graphicx}
\usepackage{grffile}
\usepackage{color}
\usepackage{afterpage}
\PassOptionsToPackage{dvipsnames}{xcolor}
\usepackage{xcolor}
\colorlet{Mycolor1}{violet!80!magenta}
\usepackage{textgreek}
\usepackage{anyfontsize}
\usepackage{hyperref}
\usepackage{xr-hyper}
\usepackage{multirow}
\usepackage[normalem]{ulem}
\usepackage{color}
\usepackage{float}
\usepackage{caption}
\usepackage{natbib}
\usepackage{bm}
\usepackage{mathpazo}
\usepackage{microtype}
\definecolor{light}{rgb}{0.5, 0.5, 0.5}

\date{\today}
\begin{document}
\title{Shear effects in active models of normal and cancer cells}
\author{Souvik Sadhukhan$^{1}$}
\author{Rajsekhar Das$^{1}$}
\author{Lin Zhao$^{2}$}
\author{Wolfgang Losert$^{2,3}$}
\author{D. Thirumalai$^{1,4}$}
\email{dave.thirumalai@gmail.com}
\affiliation{$^1$Department of Chemistry, University of Texas at Austin, Austin, Texas 78712, USA}
\affiliation{$^2$Institute for Physical Science and Technology, University of Maryland, College Park, MD, USA}
\affiliation{$^3$Department of Physics, University of Maryland, College Park, MD, USA}
\affiliation{$^4$Department of Physics, University of Texas at Austin, Austin, Texas 78712, USA}
\begin{abstract}
Mechanical properties of biological tissues, driven by passive and active forces, play a vital role in several processes ranging from development to cancer metastasis. However, the dynamical responses of cells in tissues, subject to mechanical deformations such as shear and the associated rheological properties, are not well characterized.  Here, we use three-dimensional agent-based models for normal and cancer tissues to investigate their responses to simple shear as a function of cell stiffness and stochastic active forces.  In the normal epithelium, with uniform strength of active force, the yield stress as a function of shear rate follows the Herschel-Bulkley form over a range of cell volume fraction. Strikingly,  the shear rate dependence and the elasticity-dependent changes in the yield stress fall on master curves upon suitable scaling.  To model cancer-like behavior, a certain fraction ($N_p$) of cells was chosen to have enhanced activity and decreased stiffness.  As $N_p$ increases, the extent of collective cell movement decreases, transitioning from affine (collective) to non-affine (individualistic) movement, a finding that is in accord with imaging experiments. Simulations of a model of a stiff solid tumor, with radius $R_s$ embedded in normal tissue, show that as $R_s$ increases, the yield stress increases.  Interestingly, the cells migrate collectively as $R_s$ increases. A Gaussian Mixture Model (GMM) and a mean field theory quantitatively account for the simulation as well as experimental results on cancerous, non-cancerous, and a mixture of these two  types. The combined theoretical and experimental study establishes that heterogeneity in stiffness and activity determines non-affine movements in normal and cancer tissues.
\end{abstract}
\maketitle
\vskip 1 cm

Cells in a tissue undergo large-scale mechanical deformation and rapid rearrangements during wound healing \cite{Brugues2014}, cancer metastasis \cite{Petridou2019}, gastrulation \cite{Fridtjof2024}, and embryonic development \cite{Petridou2019, Espina23JCS, Kayal2026}. These strain-rate-dependent deformations, which span diverse spatial and temporal scales, generate active stresses. For example, strain rates are low during embryogenesis, ranging from $10^{-4} s^{-1}$ to $10^{-2} s^{-1}$ \cite{Blanchard2009}, whereas in cardiac tissues they can go up to $1 s^{-1}$ to $6.5 s^{-1}$ \cite{He2005, Esfahani2021}. The generation of active mechanical stresses and diverse strain rates results in complex rheological properties that are influenced by heterogeneous mechanical microenvironments, stiffness, and activity \cite{Verdier2009}. Atomic force microscopy (AFM)  studies showed that cell stiffness and activity (self-propulsion) play a crucial role in cancer metastasis and wound healing \cite{Poujade2007, Friedrich2016, Charlotte2017, Lekka2016, Guan2021, Lee2022, Deptula2020, Fernandez2007, Dakhil2016}. For example, {individual} cancer cells are softer, more active, and fluid-like relative to their healthy counterparts \cite{Charlotte2017, Lekka2016, Guan2021}. Paradoxically, the solid tumor is stiff and has a high cell division rate, which deforms neighboring normal cells in the tissue \cite{Jain2014}. AFM experiments have shown  \cite{Deptula2020}  that cancerous human colon tissue has higher Young’s and shear modulus values than healthy tissues. These studies reveal that cellular stiffness, strain, and active stresses are affected during development and disease. 

Response to shear deformation has been used to understand the mechanical properties of a diverse range of materials~\cite{Bonn2017}, including amorphous glasses \cite{Maloney2006, Parmar2019, Thibaut2024, Lin24PNAS, Varnik2004, Ozawa2020}, granular solids \cite{Howell1999, Utter2008, Morse2021, Berthier25NatRevPhys}, foams \cite{Kabla2007}, emulsions \cite{Reboucas2025}, and active soft biological tissues \cite{Kayal2026, Sharma25NatPhys}.  {Several rheological techniques have been used to investigate the mechanical properties of materials,} such as steady shear to study flow curves and viscosity and oscillatory shear, have been used to investigate the storage and loss moduli \cite{Wyss2007}. Studies based on athermal quasistatic shear revealed the nature of plasticity and avalanche dynamics \cite{Maloney2006, Morse2021, Ruscher2021}.  Cyclic shear has been used to probe fatigue failure, memory effect, and hysteresis \cite{Leishangthem2017, Fiocco2015}. Experiments, complemented by simulations, \cite{Falk1998,Schall2007,Ridout2022,Gu2022} have reported non-affine displacement using $D^2_{min}$, which is a measure of plastic deformation. These studies yielded insights into shear transformation zones  \cite{Falk1998}, yielding transition \cite{Shrivastav2020}, energy density \cite{Leishangthem2017}, shear thinning, and shear banding \cite{Hu2023}. 

The present work is inspired by experiments  \cite{Lee2013, Gu2022}, which measured the distribution of  $D^2_{min}$ in biological tissues. In an insightful study, Gu {\it et al.} \cite{Gu2022} analyzed particle image velocimetry (PIV) using AI methods to establish that metastatic mutants of MCF10A perform individualistic motion rather than collective affine motion that characterizes non-tumor epithelial monolayers.  Cancer cells have higher $D^2_{min}$ values than the control non-tumorigenic MCF10A. Importantly, they showed that the statistical distribution of the $D^2_{min}$ is broader in metastatic cells, allowing them to conclude that $D^2_{min}$ could be used as a possible biomarker for cancer detection. However, the microscopic basis of the statistically distinct $D^2_{min}$ values between non-cancerous and cancerous cell lines is unclear.

Existing theoretical studies used two-dimensional vertex models~\cite{Popovic21NJP, Nguyen25NatComm}, focusing on the yielding transition and avalanche dynamics.  Biological tissues are composed of cells with substantial variability in size \cite{Petridou2019}, resulting in continuous division, growth, and apoptosis. In addition, the activities and stiffness of the cells are heterogeneous.   To account for these effects, we probe the response of a three-dimensional (3D) model of a polydisperse cellular system \cite{Das2024, Das2025, Kakkada2018, Guan2021} subject to constant shear as a control to assess the generic aspects of shear deformation. Using simulations of an agent-based model, we investigated the rheological properties as a function of polydispersity, elastic constants, and cell activities. The main results are:   (1) For a system with uniform activity (constant value of the {self} propulsion), the yield stress ($\sigma_{P}$) decreases with decreasing shear rate, $\dot{\gamma}$. The variation in $\sigma_{P}$ with $\dot{\gamma}$ is described by Herschel-Bulkley form \cite{Herschel1926}.  As the elasticity of cells increases, the yield stress increases as a power law with an exponent $3/4$. (2) As $\dot{\gamma}$ increases the peak height of the $P(D^2_{min})$ distribution decreases substantially. Strikingly, $P(D^2_{min}/\langle D_{min}^2 \rangle)$ collapses onto a master curve at all values of the shear rates. {The distribution of $D^2_{min}$ can be described using a probabilistic model, and two approximate limiting forms of the distribution can be derived using analytic arguments. We verified the theoretical prediction in simulations as well as on three experimental culture conditions. 
(3) To mimic cancer-like behavior, we randomly introduced a fraction of cells ($N_p$) with low stiffness and high motility. As $N_p$ increases, the yield stress decreases, and the distribution $P(D^2_{min})$ qualitatively resembles the experimental data in \cite{Gu2022} {as well as new experimental data reported in this paper.} 
(4) We also assessed the effect of introducing a solid stiff region of radius $R_s$ (emulating a solid tumor). As $R_s$ increases, the yield stress as well as the peak of $P(D^2_{min})$ increases, but the average $\langle D_{min}^2 \rangle$ decreases. 
 \section{Methods}
 \label{MethodDef}
\textit{Model:} We simulated a minimal agent-based model consisting of soft polydisperse spherical cells {in three dimensions}. Two forces govern the interaction between cells: a repulsive short-range elastic (Hertz) force and an active stochastic self-propulsion force (described below). The Hertz contact force \cite{Drasdo2005, Schaller2005, Pathmanathan2009} between  cells with radii $R_i$ and $R_j$ (measured in $\mu m$) is given by,
\begin{equation}
	F_{ij}^{el} = \frac{h_{ij}^{3/2}(t)}{\frac{3}{4} \big(\frac{1-\nu_i^2}{E_i} + \frac{1-\nu_j^2}{E_j}\big) \sqrt{\frac{1}{R_i} + \frac{1}{R_j} }},
	\label{HertzianForce}
\end{equation}
where $E_i$ (measured in MPa) and $\nu_i$ are the elastic moduli and the Poisson ratio of the cell $i$, respectively. For simplicity, we assume that $E_i$ and $\nu_i$ are the same for all cells.  The overlap between two cells  is $h_{ij}(t) = \text{max}[0, R_i + R_j - r]$  where $r = |\vec{r}_i - \vec{r}_j|$ is the center-to-center distance between the cells and $\vec{r}_i = (x_i, y_i, z_i)$ is the center of the cell $i$. 

\textit{Equation of motion:} In the overdamped limit (inertia is negligible and force is balanced by viscous drag), the dynamics of the $i^{th}$ cell is given by,
\begin{equation}
	\dot{\vec{r}}_i = \frac{\vec{F}_i}{\alpha_i} + \mu\mathcal{\vec{W}}_i(t),
\end{equation}
where the friction coefficient $\alpha_i$ is given by the Stokes-Einstein formula, $\alpha_i = 6\pi \eta^{\prime}_i R_i$, where ${\eta}^{\prime}$ is the viscosity of the medium.  The force acting on cell $i$ is  $\vec{F}_i = \sum_{j \in NN(i)} \big(F_{ij}^{el}\big) \hat{n}_{ij}$, where $NN(i)$ is the number of nearest neighbors of cell $i$. The unit vector $\hat{n}_{ij}$ acts from the center of cell $j$ to the center of cell $i$. We allowed modest uniform random variations in ${\eta}^{\prime}$ to model the inherent variability and heterogeneity of the extracellular matrix. 

The uncorrelated stochastic active force, $\mu \bm{\mathcal{W}}(t)$, where $\mu$ is the strength of the self-propulsion force (measured in $\mu m/\sqrt{s}$) is a mimic of the self-propulsion; $\bm{\mathcal{W}}$ is a Gaussian white noise with $\langle {\mathcal{W}_{i}}(t) \rangle = 0$ and correlation  $\langle \mathcal{W}^m_i(t) \mathcal{W}^n_j(t^\prime)\rangle = \delta_{mn}\delta_{ij}\delta(t-t^\prime)$.  We chose typical values of the parameters from previous studies \cite{Das2024, Das2025, Kakkada2018, Guan2021}. Polydispersity in cell sizes is measured using $\Sigma$, defined as $\Sigma = \frac{\sqrt{\langle d^2 \rangle - \langle d \rangle ^2}}{\langle d \rangle}$, where $d$ is the diameter of the cells.

We first generated equilibrated configurations close to the glass state \cite{Das2025} (see Appendix Sec.~(\ref{sec:WithourShear})). These are used as the initial conditions in the shear flow simulations. The shear deformation is implemented by adding an affine displacement, $\dot{\gamma}y_i$,  in the motion of the $x$-coordinate, where $\dot{\gamma}$ is the shear rate {(measured in $s^{-1}$)}. We used the Lees-Edwards periodic boundary conditions in the shear simulations \cite{Lees1972}. The equation of motion for all the components of the $i^{th}$ cell in the presence of  shear flow is, 
\begin{equation}
	\dot{x}_i = y_i \dot{\gamma} + \tfrac{F^x_i}{\alpha_i} + \mu \mathcal{W}^x_i(t),\;
	\dot{y}_i = \tfrac{F^y_i}{\alpha_i} + \mu \mathcal{W}^y_i(t),\;
	\dot{z}_i = \tfrac{F^z_i}{\alpha_i} + \mu \mathcal{W}^z_i(t).
	\label{EoM}
\end{equation}
The smallest timestep, $dt$, for the numerical evolution of these equations is $10s$. Using the trajectories generated in the simulations, we calculated a number of quantities to characterize the system. 

The results in the main text are reported using 1,000 cells. We also investigated the system-size dependence of the results. Except for small changes in the fitting parameters, there is no strong dependence on the system size (see Appendix Sec.~(\ref{sec:SYSSIZE})). 

{ \it Non-affine displacement, $D^2_{min}$}:
Following Falk and Langer~\cite{Falk1998}, the non-affine displacement of the cells is calculated using \( D^2_{\min} \) (measured in $\mu m^2$), which for the $j^{th}$ cell is defined as,
\begin{equation}
\label{D2minFormula}
	D^2_{\min}(t, \Delta t) = \min_{\mathcal{J} \in \mathbb{R}^{3\times 3}} \left[ \frac{1}{NN(j)} \sum_{i=1}^{NN(j)} \left\| 
	\vec{r}_i(t + \Delta t) - \vec{r}_j(t + \Delta t)
	- \mathcal{J} \left( \vec{r}_i(t) - \vec{r}_j(t) \right)
	\right\|^2 \right].
\end{equation}
In the above equation $\Delta t$ is the {time difference} between two configurations, \( NN(j) \) is the number of nearest neighbors of the cell \( j \), \( \vec{r}_i(t) \) is the position of cell \( i \) at time \( t \), and \( \mathcal{J} \) is the best-fit local affine transformation matrix that minimizes residual displacement \cite{Falk1998}. \( \mathcal{J} \) is defined as $\mathcal{J}=\mathbf Y\,\mathbf X^{-1}$, provided $\mathbf X$ is invertible. $\mathbf{X}$ and $\mathbf{Y}$ are given by, $\mathbf X = \sum_{i=1}^{NN(j)} \Delta\vec r_i(t) \otimes \Delta\vec r_i(t)$ and $\mathbf Y = \sum_{i=1}^{NN(j)} \Delta\vec r_i(t+\Delta t) \otimes \Delta\vec r_i(t)$ respectively, where $\otimes$ is the tensor product, $\Delta\vec r_i(t) = \vec r_i(t) - \vec r_j(t)$, and $\Delta\vec r_i(t+\Delta t) = \vec r_i(t+\Delta t) - \vec r_j(t+\Delta t)$. In essence, we obtain an affine deformation field,  \( \mathcal{J} \), that measures the actual displacements of the cells.  Note that $D^2_{min} \simeq 0$ for purely affine motion.  A large value of  $D^2_{min}$ ($ >> 0$) implies deviation from the affine displacement and is a measure of the non-affine or plastic motion of the cells. 

{ \it Experiments:} The non-tumorigenic human breast epithelial cell line MCF10A and GFP-expressing, tumorigenic MCF10A-derived KRas (G12V)/PTEN$^{-/-}$ cells (KP-GFP, referred to as KP cells) \cite{Lee2021} were cultured at $37^{\circ}$C and $5\%$ CO$_2$. MCF10A cells were labeled with CellTracker Orange (Thermo Fisher Scientific C34551), whereas KP-GFP cells required no additional labeling. Each cell type was resuspended at $1.5 \times 10^6$ cells/mL, and equal volumes were combined to prepare the $1:1$ mixed-cell condition. Each $6$-well plate contained six culture substrate combinations: the $1:1$ MCF10A/KP-GFP mixture on flat polycaprolactone (PCL), the mixture on nanoridged PCL, KP-GFP alone on flat PCL, KP-GFP alone on nanoridged PCL, MCF10A alone on flat PCL, and MCF10A alone on nanoridged PCL. All PCL substrates were coated with collagen IV (Corning, Cat. No. $354233$). Each culture was seeded as a 10 $\mu$L droplet and incubated overnight. Time-lapse fluorescence images were acquired using a spinning-disk confocal microscope with a $10X$ objective. Images were acquired at 3-min intervals and used to calculate $D^2_{min}$ and the $X$-component of the non-affine displacement, $x_{NA}$. We analyzed the experimental data of $D^2_{min}$ and $x_{NA}$ for MCF10A, KP, and a mixture of the two at the PCL surface. {Because the experiments come from 2D time-lapse microscopy, we used the 2D form of Eq.~(\ref{D2minFormula}) to obtain $D^2_{min}$.}

\section{Results}
\subsection {Shear-dependent mechanical response at constant activity and stiffness}
Before examining the effect of heterogeneity in the stiffness and activity, it is important to understand how the uniform (same for all the cells) activity ($\mu$) and stiffness ($E$), influence the yielding behavior and non-affine displacements as a function of the packing fraction ($\phi$). Such a model is a mimic of normal (epithelial) cells. 

\textbf{Stress overshoot and the yielding transition:}
To characterize the flow behavior, we calculated the ensemble-averaged stress, $\langle \sigma \rangle$, as a function of strain, $\gamma$, for several shear rates ranging from $10^{-4}$ to $10^{-7} s^{-1}$ (Fig.~(\ref{YieldingTrans}a)). Stress (measured in Pa) is given by
\begin{equation}
	\label{StressEq}
	\sigma_{\tilde{\alpha}\tilde{\beta}}(t) = \frac{1}{L^3} \sum_{i=1}^N\sum_{j>i}^N  r_{ij}^{\tilde{\alpha}} F_{ij}^{\tilde{\beta}},
\end{equation}
where  \( \sigma_{\tilde{\alpha}\tilde{\beta}} \) is the $\tilde{\alpha} \tilde{\beta}^{th}$ component of the stress tensor, $L$ is the system size, \( F_{ij}^\beta \) is the \( \tilde{\beta} \)-component of the force exerted by cell \( j \) on cell \( i \), and $ r_{ij}^{\tilde{\alpha}} = r_i^{\tilde{\alpha}} - r_j^{\tilde{\alpha}}$ is the \( {\tilde{\alpha}} \)-component of their separation vector. After an initial monotonic increase in stress, as the strain $\gamma$ increases, there is an overshoot (Fig.~(\ref{YieldingTrans}a)) before steady state is reached. As the shear rate increases, the yield stress increases because there is insufficient time for stress to relax through local plastic events. As a result, stress accumulates, leading to a higher yield stress ($\sigma_P$). 

We extracted $\sigma_P$ for each curve for a set of $\phi$ values to determine the dependence on  $\dot{\gamma}$ as shown in Fig.~(\ref{YieldingTrans}b). The average stress in the steady state is fit using the Herschel–Bulkley (HB) equation \cite{Herschel1926, Shrivastav2020, Lin24PNAS}, 
\begin{equation}
	\label{HBEq}
	\sigma_P(\phi,\dot{\gamma}) = \sigma_P^0(\phi) + K(\phi)\dot{\gamma}^n,
\end{equation}
where $\sigma_P^0(\phi)$, $K(\phi)$, and $n$ are the fit parameters. The value of $n < 1$ corresponds to shear-thinning behavior, which means the viscosity decreases with increasing shear rate. 
We fit the data for $\sigma_P$ in Fig.~(\ref{YieldingTrans}b) using Eq.~(\ref{HBEq}) as a function of the volume fraction, thus extending Eq.~(\ref{HBEq}) to the transient non-equilibrium situation before steady state is reached. 
The solid lines in  Fig.~(\ref{YieldingTrans}b) are the fits with $n = 0.4$ using Eq.~(\ref{HBEq}).  A plot of  $\sigma_P^0(\phi)$ as a function of $\phi$ is shown in the inset of Fig.~(\ref{YieldingTrans}b). After rearrangement,  Eq.~(\ref{HBEq}) becomes,
\begin{equation}
	\label{HBEqCollapse}
	\frac{\sigma_P(\phi,\dot{\gamma})}{\sigma_P^0(\phi)} = 1 + (\tau_c\dot{\gamma})^{0.4},
\end{equation}
where $\tau_c = \big(\frac{K(\phi)}{\sigma_P^0(\phi)}\big)^{1/0.4}$ is a density-dependent timescale. The above equation suggests that  $\frac{\sigma_P(\phi,\dot{\gamma})}{\sigma_P^0(\phi)}$  versus  $\tau_c\dot{\gamma}$ should result in data collapse because the right hand side of Eq.~(\ref{HBEqCollapse}) is independent of $\dot{\gamma}$ and $\phi$. The solid line in Fig.~(\ref{YieldingTrans}c) confirms that this is indeed the case. Fig.~(\ref{YieldingTrans}d) shows that $\sigma_P(\phi,\dot{\gamma})$ varies linearly with $\phi$ for a fixed $\dot{\gamma}$ and can be fit using, 
\begin{equation}
	\label{HBEqvariationPhi}
	\sigma_P(\phi,\dot{\gamma}) = P(\dot{\gamma}) + \phi Q(\dot{\gamma}),
\end{equation}
where $P(\dot{\gamma})$ and $Q(\dot{\gamma})$ are two  parameters. The symbols in Fig.~(\ref{YieldingTrans}d) are the simulation data, and the solid lines are the fits using Eq.~(\ref{HBEqvariationPhi}). We plotted the variation of the two fitting parameters  $P(\dot{\gamma})$ and $Q(\dot{\gamma})$  as a function of $\dot{\gamma}$ in Fig.~(\ref{YieldingTrans}e) and the inset, respectively. Plot of $\frac{\sigma_{P}(\phi, \dot{\gamma})-P(\dot{\gamma}))}{Q(\dot{\gamma})}$ as a function of $\phi$, shows that the data fall a single master curve (Fig.~(\ref{YieldingTrans}f)). 

To assess how the system parameters affect the yielding transition, we varied the self-propulsion, $\mu$, as shown in Fig.~(\ref{YieldingTrans}g). We explored the effect of $\mu$ without shear in Appendix Sec.~(\ref{sec:WithourShear}). As $\mu$ decreases, the yield stress increases and finally saturates, as shown in Fig.~(\ref{YieldingTrans}h).  {Upon increasing $E$, the yield stress increases (Fig.~(\ref{YieldingTrans}i)).}   The plot of $\sigma_P(E)$ as a function of $E$ for various $\phi$ in Fig.~(\ref{YieldingTrans}j) shows a power law increase, $f(E) = aE^\beta$, with $\beta = 3/4$. 
The inset in Fig.~(\ref{YieldingTrans}j) shows that $\sigma_P(E)/a$ as a function of $E^{3/4}$ results in the collapse of the data into a single line. Interestingly, polydispersity, $\Sigma$, does not show a significant change in the yield stress (see Appendix Sec.~(\ref{sec:EffectofPoly})).

\textbf{Non-affine displacements:}
To measure the non-affine displacements caused by local shear deformation with respect to the neighbors, we used the Falk-Langer \cite{Falk1998} method. The intensity plots of $D^2_{min}(t, \Delta t)$ in the sheared configuration at strain, $\gamma = 0.2$, for $3$ different values of $\Delta t$ are shown in Fig.~(\ref{D2minProps}a-c). For small $\Delta t$, the rearrangements are small and localized. However, as $\Delta t$ increases, the size of the rearrangements increases. The distribution of $D^2_{min}$ as a function of $\Delta t$ is discussed in Appendix Sec.~(\ref{sec:EffectofPoly}). Next, we calculated  $\langle D^2_{min} \rangle \equiv
\langle D^2_{min}(0, \Delta t) \rangle$, measured with respect to the unstrained state ($\gamma = 0$),  for a range of $\dot{\gamma} = 10^{-6}-10^{-4} s^{-1}$ (Fig.~(\ref{D2minProps}d)). For high $\dot{\gamma}$, the average $\langle D^2_{min} \rangle$ is small and for small $\dot{\gamma}$ the average $\langle D^2_{min} \rangle$ is large. For small $\dot{\gamma}$, there is ample time for stress relaxation through highly non-affine local plastic rearrangements, leading to large $D^2_{min}$ values. On the other hand, at high $\dot{\gamma}$, affine deformation dominates and the particles are forced to follow the imposed shear field more coherently, leaving less time for localized relaxations, resulting in small $D^2_{min}$ values. 

To quantify relaxation dynamics, we calculated the {two-point} self-overlap function, $Q(\gamma)$,\cite{Guo1995, Tang2023},
\begin{equation}
	\label{QgammaEq}
	Q(\gamma) =\langle\frac{1}{N} \sum_{i= 1}^N W(D^2_{min}) \rangle,
\end{equation}
where $N$ is the number of cells, $W$ is the Heaviside step function, $W(D^2_{min}) = 1$ if $D^2_{min} \leq D_c^2$ and $W(D^2_{min}) = 0$ if $D^2_{min} > D_c^2$. We choose $D_c^2 = 20 \mu m^2$, which is around the plateau of $\langle D^2_{min} \rangle$ (see Fig.~(\ref{D2minProps}d)). Fig.~(\ref{D2minProps}e) shows that $Q(\gamma)$ decays faster at low $\dot{\gamma}$ and slower for high $\dot{\gamma}$. 

As a measure of dynamic heterogeneity, we also computed the fluctuations in $Q(\gamma)$ using the four-point dynamic susceptibility introduced in the context of spin glasses and structural glasses, $\chi_4(\gamma)$~\cite{Kirkpatrick1988}. We calculated $\chi_4(\gamma)$ using the fluctuations in the overlap function~\cite{Guo1995, Tang2023}, 
\begin{equation}
	\label{chi4gammaEq}
	\chi_4(\gamma) =N[\langle Q(\gamma)^2 \rangle - \langle Q(\gamma) \rangle^2].
\end{equation}
A plot of $\chi_4(\gamma)$ (Fig.~(\ref{D2minProps}f)) shows that the fluctuations are much larger at small $\dot{\gamma}$ compared to high $\dot{\gamma}$.  For small $\dot{\gamma}$, the cells are heterogeneous and exhibit more variability in the $D^2_{min}$ values. We extracted the strain values at which $\chi_4(\gamma)$ reaches a maximum and similarly for the stress, $\sigma$ (see Fig.~(\ref{YieldingTrans}a)). The dependence of these quantities as a function of $\dot{\gamma}$ is shown in Fig.~(\ref{D2minProps}g). The curves tend to saturate for large $\dot{\gamma}$. When plotted against each other, there is a linear relationship, implying that the maximum heterogeneity in non-affine displacement and strain value at the yield stress is correlated (Inset of Fig.~(\ref{D2minProps}g)). 

To determine how the distribution changes at various stages of yielding, we calculated $D^2_{min}(0, \Delta t)$ at five values of $\gamma$ (Fig.~(\ref{D2minProps}h)).  The width of the $P(D^2_{min})$ increases as $\gamma$ increases, implying the presence of larger non-affine displacements. The scaled distribution $D^2_{min}/\langle D^2_{min} \rangle$, the curves collapse into a single master curve as shown in Fig.~(\ref{D2minProps}i), implying that universal statistics govern non-affine rearrangements.


\subsection{Approximate functional form of \texorpdfstring{$P(D^2_{min})$}{PD2min}}
To obtain analytic insights, we used the Gaussian Mixture Model (GMM), which is based on a clustering algorithm \cite{Dempster1977}, to fit $P(D^2_{min})$.  The GMM represents the distribution as a mixture of $K$ Gaussian terms as,
   \begin{equation}
   	\label{GMM}
   	p(\mathbf{x}) = \sum_{k=1}^K \pi_k \mathcal{N}(\mathbf{x}I \mu_k,\Sigma_k),
   \end{equation}
   where $\pi_k$ is the prior probability of the $k$-th Gaussian. $\sum_{i = 1}^K \pi_k = 1$ and $0\leq \pi_k \leq 1$. $\mathcal{N}$ is a multivariate Gaussian distribution with unknown parameters $\mu_k$ and $\Sigma_k$. We refer GMM-fit as the fits of $P(D^2_{min})$  in log-transformed $D^2_{min}$ values with $K = 2$. The fits  of $P(D^2_{min})$  for varying values of $\gamma$ and $\mu$ in (see Fig.~(\ref{FunctionalForms}a) and Fig.~(\ref{FunctionalForms}b)) show that GMM captures the full distribution of $D^2_{min}$. 
	
To uncover the underlying distribution of $D^2_{min}$, we followed the approach used by Utter and Behringer \cite{Utter2008} using the non-affine displacements, $\delta r$, of the individual cells. Instead of assuming that the displacement distributions are Gaussian, we took them as non-Gaussian, $P(\delta r_i) \simeq A \exp(-B|\delta r_i|^\alpha)$ with $\alpha \leq 2$, where $A, B$ are constants. We are interested in deriving the limiting form of the distribution, using extreme value statistics \cite{Gumbel2013} (see Appendix Sec.~(\ref{sec:MFtheo}) for details). The limiting forms for two extremes are given by,
\begin{equation}
	P(D^2_{min})=
	\begin{cases}
		C_1({D^2_{min}})^{N_g/2 -1},& \text{for  } D^2_{min} \rightarrow 0 \\
		C_2 \exp[-D(D^2_{min})^{\alpha/2}],& \text{for  } D^2_{min} \rightarrow \infty,
	\end{cases}
	\label{D2minTheo}
\end{equation}
{where $C_1, N_g, C_2, D$, and $\alpha$ are fitting parameters.}  
The exponent of $D^2_{min}$ within the exponential in the large $D^2_{min}$ expression is exactly half the value of the non-Gaussian exponent. 

To assess the accuracy of the theoretical formula, we fit the distribution of the $X$-component of the non-affine displacement with this form $P(x_{NA}) = A_1 \exp(-B_1|x_{NA}|^\alpha)$, and obtained the value of $\alpha$. Then we used the same $\alpha$ for fitting the $D^2_{min} \rightarrow \infty$ form. For our data, we assumed that $N_g$ is a constant in Eq.~(\ref{D2minTheo}) because it does not vary much. With this simplification, there is one fitting parameter, $C_1$, for the small $D^2_{min}$ part and two fitting parameters, $C_2$ and $D$, for the large $D^2_{min}$ part. We calculated the non-affine displacement distribution and the distribution of the $D^2_{min}$ for $\Delta t = 100$ s at $\gamma = 0.2$ (close to the yielding point) for various $E$ values, as shown in Fig.~(\ref{FunctionalForms}c) and Fig.~(\ref{FunctionalForms}d). The value of  $\alpha$ varied between $\sim 1.86-1.46$ as shown in Fig.~(\ref{FunctionalForms}c). The $P(D^2_{min})$ for the stiffer cell system (high $E$) is sharp, whereas the soft system (low $E$) has a much broader distribution, as shown in Fig.~(\ref{FunctionalForms}d). The effect of polydispersity in $P(D^2_{min})$ in Appendix Sec.~(\ref{sec:EffectofPoly}). 
For lower $E$,  the system relaxes like a liquid with larger rearrangements, making $P(D^2_{min})$ broader. 

We have obtained the $P(D^2_{min})$ for three different sets of experiments and fitted each set with the GMM model. The experimental data agree well with the GMM prediction as shown in Fig.~(\ref{FunctionalForms}e). To compare our mean-field expression, of $P(D^2_{min})$, we computed the distribution of the $X$-component of the non-affine displacement with this form $P(x_{NA})$, and fitted it with the form $P(x_{NA}) = A_1 \exp(-B_1|x_{NA}|^\alpha)$ (Fig.~(\ref{FunctionalForms}f)). We found that $\alpha$ varies between $\sim 1.05-1.15$. We used the same $\alpha$ for fitting the $D^2_{min} \rightarrow \infty$ form as shown in Fig.~(\ref{FunctionalForms}g) and (h) on a semi-log scale and normal scale, respectively. The fittings with our mean-field theory were restricted to the well-sampled $D^2_{min} \rightarrow \infty$ limit, and the $D^2_{min} \rightarrow 0$ limit was excluded because of limited sampling. These results suggest the diverse applicability of our theory in experiments on {three experimental culture conditions}: cancerous, non-cancerous, and a mixture of the two.

\subsection{Activity and stiffness variations effects on non-affine displacements}

The results in the previous section set the stage for calculating $D_{min}^2$ in models of cancer-like behavior.  To assess if the simulations could provide insights into experiments that have measured the $D^2_{min}$ distribution in cancer cells~\cite{Gu2022}, we generalized the model. Minimally, a physical model for cancer should account for two cell types with different stiffness and self-propulsion. 
Experimental studies \cite{Fuhs2022, Charlotte2017} suggest that, although the tumor is stiff and less dynamic, the individual cancer cells are soft and more dynamic. The invading ability of the cells through the tissue is controlled by the cell stiffness \cite{Gajda2025}. During metastasis, the cancerous cells become motile and migrate to distant sites after rupturing from their primary sites.   
In the presence of two types of cells, there could be a competition between the dynamic soft cells and the stiff, slow-moving cells. We expect such an interplay to affect the yielding as well as the non-affine displacement distribution. To connect to the measured~\cite{Gu2022} non-affine displacements, we considered three models with cells having different activities and stiffness.  

\textbf{Mechanical response of the cellular system through random cell softening:} The first model is based on the observation that individual cancer cells have lower stiffness and higher deformability than the non-cancerous counterparts \cite{Suresh2007, Lee2012}. To probe the effect of cell softness in a heterogeneous environment, $N_p$ randomly chosen cells were made more dynamic and softer compared to the rest of the cells by altering their $\mu = 0.07 \mu m/\sqrt{s}$ and $E = 0.5E_0$, where $E_0 = 0.001$ MPa \cite{Gnan2019, Kakkada2018, Das2024}. (Fig.~(\ref{TestSphereSimRandomNp}a)). The rest of the cells have $\mu = 0.045\mu m/\sqrt{s}$ and $E = 5E_0$. As the number of soft dynamic cells increases, the yield stress decreases  (Fig.~(\ref{TestSphereSimRandomNp}b)) because the system becomes fluid-like. To qualitatively compare our findings with experiments \cite{Gu2022}, we extracted the data using the software WebPlotDigitizer \cite{WebPlotDigitizer} and displayed them as a bar plot in Fig.~(\ref{TestSphereSimRandomNp}c). The average $D^2_{min}$ increases in going from a non-tumorigenic or dormant variant of MCF10A to a tumorigenic or metastatic KRas and KRas/PTEN$^{-/-}$. It has a maximum value for a highly aggressive, invasive human breast cancer cell line, MDA-MB-231. The calculated  $D^2_{min}$ as a function of $N_p$ shows that the average $D^2_{min}$ increases monotonically with $N_p$ (see right y-axis of the same plot), which is in qualitative agreement with experiments \cite{Gu2022, Lee2012}. 
 
To confirm that increasing $N_p$ mimics more metastatic cancerous cells, we calculated $P(D^2_{min})$ as a function of $N_p$ (Fig.~(\ref{TestSphereSimRandomNp}d)). Gu {\it et al.} \cite{Gu2022} reported that the peak height of $P(D^2_{min})$ decreases, and the distribution is shifted towards the right, from non-tumorigenic to metastatic cells.  Our simulations show that as $N_p$ increases, the peak of $P(D^2_{min})$ decreases because the dynamics becomes more heterogeneous and non-affine, which accords well with experiments.  The randomly chosen $N_p$ cells undergo more individualistic motion resembling the metastatic cells, resulting in a high mean $D^2_{min}$ and a broader tail in the $P(D^2_{min})$ distribution. We conclude that the dissimilarity in the stiffness is one of the causes of the changes in the $D^2_{min}$ values. By randomly making \(N_p\) cells very soft and more active than the rest of the cells, there is a competition among activity, elastic caging, resistance, force transmission, and the deformability of the two types of cells. The propensity of soft active cells to escape the elastic cage formed by the stiff cells depends on whether the strength of the active force is sufficiently large to overcome the local Hertz restoring forces. Once the soft active cells manage to surpass this barrier through stochastic active forces, plastic rearrangement occurs. Increasing \(N_p\) injects stronger stochastic active forces, leading to significantly larger local overlaps and rearrangements. Given the size of these rearrangements, the \(D^2_{min}\) values increase, which shifts and broadens the peak of \(P(D^2_{min})\). Furthermore, the force between two stiff cells is considerably stronger than the forces between two soft cells or between one soft and one stiff cell. Consequently, an increase in \(N_p\) softens and weakens the force networks, causing the system to yield more easily and become increasingly non-affine.

To understand how the activities of the randomly chosen soft cells and the rest of the cells affect the yielding and $P(D^2_{min})$, we constructed 3D and 2D phase space plots of $\mu_{N_p}-\mu_{rest}-\sigma_P$ (Fig.~(\ref{TestSphereSimRandomNp}e-f)) and $\mu_{N_p}-\mu_{rest}-\text{Maxval of }P(D^2_{min})$ (Fig.~(\ref{TestSphereSimRandomNp}g-h)). Here,  $\mu_{N_p}$ is the activity of the randomly chosen $N_p$ cells, $\mu_{rest}$ is the activity of the rest of the cells, $\sigma_{P}$ is the yield stress value, and $\text{Maxval of }P(D^2_{min})$ is the peak value of the $P(D^2_{min})$. The activity of the $N_p$ cells satisfies $\mu_{N_p}>\mu_{rest}$.  Fig.~(\ref{TestSphereSimRandomNp}f) and Fig.~(\ref{TestSphereSimRandomNp}h) are the corresponding 2D representation of Fig.~(\ref{TestSphereSimRandomNp}e) and Fig.~(\ref{TestSphereSimRandomNp}g) respectively. 
{The colorbar in Fig.~(\ref{TestSphereSimRandomNp}e-f) is the yield stress value, $\sigma_P$.  The $\text{Maxval of }P(D^2_{min})$ is given in Fig.~(\ref{TestSphereSimRandomNp}g-h).}  We interpolated the phase space for better visualization. The phase space plots of $\mu_{N_p}-\mu_{rest}-\sigma_P$ and $\mu_{N_p}-\mu_{rest}-\text{Maxval of }P(D^2_{min})$ are similar.  For a fixed $\mu_{rest}$ and increasing $\mu_{N_p}$, the yield stress decreases, as does the peak of  $P(D^2_{min})$ because the cells become fluid-like due to the higher activity of the $N_p$ cells. 

To understand the phase space diagram, we calculated the $P(D^2_{min})$ for all the cells, the $N_p$ cells, and the rest of the $(N-N_p)$ cells for four activity values (Fig.~(\ref{TestSphereSimRandomNp}i-l) for $N_p = 250$). Because the rest of the cells are stiffer and less active compared to the $N_p$ cells, the peak of $P(D^2_{min})$ is the highest for those cells.  The distribution $P(D^2_{min})$ for the full sample is between the $P(D^2_{min})$s of $N_p$ cells and the rest of the cells. The $P(D^2_{min})$ for the full sample is close to the $P(D^2_{min})$ of {the rest of the cells.}  When both activities are low and equal $(\mu_{N_p} = \mu_{rest} = 0.045 \mu m/\sqrt{s})$, 
it is the stiffness difference that plays the key role. As the activities are low and the number of stiff cells is higher, it leads to an increase in the peak of $P(D^2_{min})$. So, for this set of parameters, the peak of $P(D^2_{min})$ is high, as shown in Fig.~(\ref{TestSphereSimRandomNp}i). 

A different scenario emerges in  Fig.~(\ref{TestSphereSimRandomNp}l), where three contributions decrease the peak of $P(D^2_{min})$. Because in this case the activities, $\mu_{N_p} = 0.07 \mu m/\sqrt{s}$ and $\mu_{rest} = 0.07 \mu m/\sqrt{s}$, are high, the peak of $P(D^2_{min})$ decreases. The $N_p$ cells have low stiffness values ($0.5E_0$), which also contribute to the decrease of the peak. Only the $(N - N_p)$ stiff cells contribute to the increase in the peak of $P(D^2_{min})$. The net result is that the peak of $P(D^2_{min})$ is low, as shown in Fig.~(\ref{TestSphereSimRandomNp} l). 

The situation becomes complicated when the activity values ($\mu$) differ. In this case, there is competition between the cell stiffness and activities, which determines the fate of the peak of $P(D^2_{min})$ (Fig.~(\ref{TestSphereSimRandomNp} j) and (k)). These correspond to the intermediate color gradient regime of the phase space diagram. Hence, the phase space plots give insights into the change in yield stress and the peak of  $P(D^2_{min})$ due to the underlying heterogeneity in the activities. We compare our findings with experiments by displaying $P(D^2_{min})$ versus $D^2_{min}$ in Fig.~(\ref{TestSphereSimRandomNp} m) for a mixture of cancerous and non-cancerous cells, cancerous KP cells, and non-cancerous MCF10A cells. The three cases mimic all the cells, $N_p$ cells, and the  cells other than the $N_p$ cancer cells, respectively, in the simulations. The plot shows that the experimental data qualitatively agree with  with the simulation results, which implies heterogeneity in the activities and stiffness in the cell lines control the behavior of $P(D^2_{min})$.

\textbf{Effect of spherical region of stiff cells embedded in a sea of normal cells:} The measured values of the stiffness of normal mammary gland tissue range from $0.1-0.2 E_0$, whereas malignant mammary tumors have high stiffness of  $1-4 E_0$, where $E_0 = 0.001MPa$ \cite{Paszek2005, Levental2009}. To account for these differences, we created a second model mimicking the solid stiff tumor, which was created by embedding a spherical region of radius $R_s$ inside the normal cells. The cells inside the spherical region are stiff with $E = 5E_0$ (red cells in Fig.~(\ref{TestSphereSimSphRadius}a)). The value of cell elasticity for the rest of the cells is $E = 0.5E_0$ (blue cells in Fig.~(\ref{TestSphereSimSphRadius}a)).  The value of $\mu$ for all the cells is the same ($0.045 \mu m/\sqrt{s}$) in Fig.~(\ref{TestSphereSimSphRadius}a-d).  As  $R_s$ increases, there is an increase in the number of stiff cells, emulating a growing tumor.  

The ensemble-averaged stress versus strain as a function of $R_s$ in Fig.~(\ref{TestSphereSimSphRadius}b) shows that as the size of the solid tumor increases, the yield stress value increases. Next, we calculated  $P(D^2_{min})$ separately for the stiff cells surrounding the rest of the soft cells, and for all the cells for a fixed $R_s = 70 \mu m$ (Fig.~(\ref{TestSphereSimSphRadius}c)). $P(D^2_{min})$ for $R_s = 0$ (all the cell have uniform $E$) is also shown as a reference.   The inclusion of stiff cells increases the peak of the distribution of $D^2_{min}$ locally for both the stiff as well as the soft surrounding cells, which implies that the solid stiff tumor also affects the dynamics of its neighboring cells. In Fig.~(\ref{TestSphereSimSphRadius}d), we  plot $P(D^2_{min})$ versus $D^2_{min}$ for various $R_s$. As $R_s$ increases, the system tends toward solid-like affine states, resulting in less variability in $D^2_{min}$. This explains the increase in the peak height in $P(D^2_{min})$  with increasing $R_s$.  Mean $D^2_{min}$ decreases as $R_s$  increases, as shown in the inset of Fig.~(\ref{TestSphereSimSphRadius}d).

To understand the effect of the stiff spherical region, we varied the $\mu$ values. Let  $\mu$ of cells within the spherical region be $\mu_{core}$ and for the surrounding cells be $\mu_{surr}$. We  constructed 3D and 2D phase spaces of $\mu_{core}-\mu_{surr}-\sigma_P$ (Fig.~(\ref{TestSphereSimSphRadius}e-f)) and $\mu_{core}-\mu_{surr}-\text{Maxval of }P(D^2_{min})$ (Fig.~(\ref{TestSphereSimSphRadius}g-h)). Fixing $\mu_{surr}$ and increasing $\mu_{core}$ makes the cellular system more heterogeneous and decreases the yield stress as well as $\text{Maxval of }P(D^2_{min})$. It is clear that heterogeneity in the activities and stiffness affects the statistical distribution of $D^2_{min}$ as well as the mechanical properties of the cells.

\textbf{Effect of introducing a spherical region of slow cells on mechanical properties:} In the third model, we made the cells within the spherical region less dynamic ($\mu = 0.045 \mu m/\sqrt{s}$) than the surrounding cells ($\mu = 0.065 \mu m/\sqrt{s}$) while keeping $E$ constant for all cells.  The ensemble-averaged stress versus strain for four values of $R_s$ in Fig.~(\ref{TestSphereSimChangingMu}a) shows an increase in the yield stress.  The peak of the $D^2_{min}$ distribution also increases slightly, as shown in Fig.~(\ref{TestSphereSimChangingMu}b). Taken together, these results show that the change in the peak of $P(D^2_{min})$ and the yield stress is more prominent when stiffness is changed compared to activity.

There is a competition between the cells with different types of activities and stiffness. In reality, the cells are polydisperse with variations in both stiffness and activity. The phase space plots show that non-affine displacements are determined by the competition between the cells with varying characteristics. When we introduce a stiff spherical core inside the soft active cellular environment, there is a competition between cells with dissimilar properties. The stiff cellular core resists deformation, stores elastic stress, and moves more affinely under shear. These create a strong mechanical heterogeneity in the cellular system by preventing local rearrangements. At the interface between the soft and stiff regions, stiff cells and soft cells respond differently even if the activities are similar. The stiff cells resist deformation, while the soft cells deform readily, which creates a mismatch in their displacements. 
If the stiff cells are more active, then there are two competing effects. On the one hand, higher stiffness suppresses local rearrangement.  On the other, higher activity promotes rearrangement. The stiff core resists yielding, but the high activity drives rearrangements at the interface of soft surroundings and stiff core. Each region contributes differently due to these competing effects to yielding and $P(D^2_{min})$, which are reflected by the gradient of colors in the phase space plots.

\section{DISCUSSION AND CONCLUSIONS}
In this combined theoretical and experimental study, we investigated the rheological properties of three-dimensional polydisperse cells subject to simple shear. The simulations were done using an agent-based model as a function of activity and cell elasticity.  Remarkably, without tuning any parameter in the simulations, we found qualitative similarities for $P(D^2_{min})$ with  {the experimental data on three diverse classes of experimental culture conditions: cancerous, non-cancerous and mixture of the two.} The simulations allow us to connect cell rheology and metastasis using non-affine displacements.  In the process, we provide qualitative insights into experiments \cite{Gu2022}. Our study reveals that heterogeneity in self-propulsion and stiffness are the key factors that control the change in the non-affine displacement distribution during metastasis. 

This work is broadly divided into two parts. In the first part, we characterized the mechanical transient properties of a system with fixed stiffness and activity. The yield stress as a function of the shear rates obeys a Herschel–Bulkley (HB) type of behavior. We  investigated the transient yielding characteristics of the cells under  ``physiological" conditions by varying the packing fractions ($\phi$), shear rates ($\dot{\gamma}$), activity ($\mu$), elasticity of cells ($E$), and polydispersity $(\Sigma)$ of the cells.   The calculated distribution of the affine displacement parameter as a function of the strain falls on a single curve when scaled by the mean value of $\langle D_{min}^2 \rangle$.   The dynamics associated with the plastic events were quantified with the $Q(\gamma)$ and the heterogeneity using $\chi_4(\gamma)$. It is found that $Q(\gamma)$ decays faster at low $\dot{\gamma}$ and slower for high $\dot{\gamma}$.  The reason is that the $\langle D^2_{min}\rangle$ values are low at high $\dot{\gamma}$ because the cells flow mostly affinely. The $\chi_4(\gamma)$ peak increases with decreasing $\dot{\gamma}$ because at low shear rates the $D^2_{min}$ values are more variable and non-affine. 

To obtain analytic insights, we used a probabilistic Gaussian mixture model to fit the distribution of $D^2_{min}$ in log-transformed $D^2_{min}$ values. Using extreme value statistics and a few approximations \cite{Utter2008},  we derived the limiting form for $P(D^2_{min})$.  The growth of $P(D^2_{min})$ exhibits a power law, and the decay at large $D_{min}^2$  is best described by a stretched exponential decay. The $X$-component of the non-affine displacement also follows the stretched exponential.  The exponent of the former is related to the latter by a factor of two. The theory successfully  explains the simulation results and experimental findings under {three different culture conditions.}

In the second part, we calculated  $P(D^2_{min})$ by varying activity and stiffness to mimic the effect of cancer cells. These simulations consider two types of cells.  The yield stress for randomly chosen $N_p$ cells with very low stiffness and high activity compared to the rest of the cells in the system decreases as $N_p$ increases.  {The distributions $P(D^2_{min})$}  become broader with the shift to higher values, which is in qualitative accord with experiments \cite{Gu2022}. The mean $D^2_{min}$ increases with $N_p$ because the system becomes more heterogeneous. 
From the phase space plots, we infer that the cellular system becomes more non-affine and fluid-like upon an increase in $\mu_{N_p}$ for a fixed $\mu_{rest}$. Qualitatively similar   trends for  $P(D^2_{min})$ are found in experiments for {three different experimental culture conditions: cancerous, non-cancerous and a mixture of these two.}

We also investigated the presence of a spherical region of radius $R_s$ embedded in a sea of normal cells. The elasticity of the cells $E$ is large compared to the cells outside the region.  As the radius $R_s$ increases, the peak of $P(D^2_{min})$  as well as the yield stress subject to shear increases. Introducing such a region affects not only global tissue properties, but also the neighboring tissues. We also calculated the 2D and 3D phase space plots as a function of $\mu_{core}-\mu_{surr}-\sigma_P$ and $\mu_{core}-\mu_{surr}-\text{Maxval of }P(D^2_{min})$.  Taken together, our results show that the heterogeneity in stiffness and activity is the key reason for the emergence of non-affinity in models of cancer tissues.


\textbf{Acknowledgments:}   This research was supported by the National Science Foundation (PHY 2310639) and the Collie-Welch Chair through the Welch Foundation (F-0019). {WL and LZ acknowledge funding from the POLS student research network grant UMD-NSF PHY2310742.  The experimental data used in this work were collected using the Nikon W-1 (S10 OD026698).} 

\bibliographystyle{unsrt}
\bibliography{citationsFinal}

@Article{Lee2021,
author={Lee, Rachel M.
and Vitolo, Michele I.
and Losert, Wolfgang
and Martin, Stuart S.},
title={Distinct roles of tumor associated mutations in collective cell migration},
journal={Scientific Reports},
year={2021},
month={May},
day={13},
volume={11},
number={1},
pages={10291},
issn={2045-2322},
doi={10.1038/s41598-021-89130-6},
}

@article{Schall2007,
author = {Peter Schall  and David A. Weitz  and Frans Spaepen },
title = {Structural Rearrangements That Govern Flow in Colloidal Glasses},
journal = {Science},
volume = {318},
number = {5858},
pages = {1895-1899},
year = {2007},
doi = {10.1126/science.1149308}}

@article{Ridout2022,
author = {Sean A. Ridout  and Jason W. Rocks  and Andrea J. Liu },
title = {Correlation of plastic events with local structure in jammed packings across spatial dimensions},
journal = {Proceedings of the National Academy of Sciences},
volume = {119},
number = {16},
pages = {e2119006119},
year = {2022},
doi = {10.1073/pnas.2119006119}}

@article{Brugues2014,
	author="Agust{\'{i}} Brugu{\'{e}}s and Ester Anon and Vito Conte and Jim H. Veldhuis and Mukund Gupta and Julien Colombelli and Jos{\'{e}} J. Mu{\~{n}}oz and G. Wayne Brodland and Benoit Ladoux and Xavier Trepat",
	title="Forces driving epithelial wound healing",
	journal="Nat. Phys.",
	volume="10",
	pages="683",
	year="2014",
	doi="10.1038/NPHYS3040",
}

@article{Bonn2017,
  title = {Yield stress materials in soft condensed matter},
  author = {Bonn, Daniel and Denn, Morton M. and Berthier, Ludovic and Divoux, Thibaut and Manneville, S\'ebastien},
  journal = {Rev. Mod. Phys.},
  volume = {89},
  issue = {3},
  pages = {035005},
  numpages = {40},
  year = {2017},
  month = {Aug},
  publisher = {American Physical Society},
  doi = {10.1103/RevModPhys.89.035005},
}

@misc{WebPlotDigitizer,
    author = {Ankit Rohatgi},
    title = {WebPlotDigitizer},
    url = {https://automeris.io},
    version = {4},
}

@article{Poujade2007,
author = {M. Poujade  and E. Grasland-Mongrain  and A. Hertzog  and J. Jouanneau  and P. Chavrier  and B. Ladoux  and A. Buguin  and P. Silberzan },
title = {Collective migration of an epithelial monolayer in response to a model wound},
journal = {Proceedings of the National Academy of Sciences},
volume = {104},
number = {41},
pages = {15988-15993},
year = {2007},
doi = {10.1073/pnas.0705062104}}

@article{Esfahani2021,
author = {Amir Monemian Esfahani  and Jordan Rosenbohm  and Bahareh Tajvidi Safa  and Nickolay V. Lavrik  and Grayson Minnick  and Quan Zhou  and Fang Kong  and Xiaowei Jin  and Eunju Kim  and Ying Liu  and Yongfeng Lu  and Jung Yul Lim  and James K. Wahl  and Ming Dao  and Changjin Huang  and Ruiguo Yang },
title = {Characterization of the strain-rate–dependent mechanical response of single cell–cell junctions},
journal = {Proceedings of the National Academy of Sciences},
volume = {118},
number = {7},
pages = {e2019347118},
year = {2021},
doi = {10.1073/pnas.2019347118}}

@article{He2005,
  author    = {He, Z. and Ritchie, J. and Grashow, J. S. and Sacks, M. S. and Yoganathan, A. P.},
  title     = {In vitro dynamic strain behavior of the mitral valve posterior leaflet},
  journal   = {Journal of Biomechanical Engineering},
  volume    = {127},
  number    = {3},
  pages     = {504--511},
  year      = {2005}
}

@article{Blanchard2009,
  author    = {Blanchard, G. B. and Kabla, A. J. and Schultz, N. L. and Butler, L. C. and Sanson, B. and Gorfinkiel, N. and Mahadevan, L. and Adams, R. J.},
  title     = {Tissue tectonics: morphogenetic strain rates, cell shape change and intercalation},
  journal   = {Nature Methods},
  volume    = {6},
  number    = {6},
  pages     = {458--464},
  year      = {2009},
  doi       = {10.1038/nmeth.1327},
}

@inbook{Parmar2019,
author = {Anshul D. S. Parmar and Srikanth Sastry},
title = {Mechanical Behaviour of Glasses and Amorphous Materials},
booktitle = {Advances in the Chemistry and Physics of Materials},
publisher = {WORLD SCIENTIFIC},
chapter = {Chapter 21},
pages = {503-527},
doi = {10.1142/9789811211331_0021},
year  ={2019}
}

@Article{Gajda2025,
author={Gajda, Alexa M.
and Rodr{\'i}guez-L{\'o}pez, Raymundo
and Er, Ekrem Emrah},
title={Targeting cancer cell stiffness and metastasis with clinical therapeutics},
journal={Clinical {\&} Experimental Metastasis},
year={2025},
month={Jun},
day={11},
volume={42},
number={4},
pages={34},
issn={1573-7276},
doi={10.1007/s10585-025-10353-2},
}

@article{Suresh2007,
	title = {Biomechanics and biophysics of cancer cells},
	journal = {Acta Biomaterialia},
	volume = {3},
	number = {4},
	pages = {413-438},
	year = {2007},
	issn = {1742-7061},
	doi = {https://doi.org/10.1016/j.actbio.2007.04.002},
	author = {Subra Suresh}
}

@article{Lee2012,
	title = {Mismatch in Mechanical and Adhesive Properties Induces Pulsating Cancer Cell Migration in Epithelial Monolayer},
	journal = {Biophysical Journal},
	volume = {102},
	number = {12},
	pages = {2731-2741},
	year = {2012},
	issn = {0006-3495},
	doi = {https://doi.org/10.1016/j.bpj.2012.05.005},
	author = {Meng-Horng Lee and Pei-Hsun Wu and Jack Rory Staunton and Robert Ros and Gregory D. Longmore and Denis Wirtz}
}

@article{Levental2009,
	title = {Matrix Crosslinking Forces Tumor Progression by Enhancing Integrin Signaling},
	journal = {Cell},
	volume = {139},
	number = {5},
	pages = {891-906},
	year = {2009},
	issn = {0092-8674},
	doi = {https://doi.org/10.1016/j.cell.2009.10.027},
	author = {Kandice R. Levental and Hongmei Yu and Laura Kass and Johnathon N. Lakins and Mikala Egeblad and Janine T. Erler and Sheri F.T. Fong and Katalin Csiszar and Amato Giaccia and Wolfgang Weninger and Mitsuo Yamauchi and David L. Gasser and Valerie M. Weaver}
}

@article{Paszek2005,
	title = {Tensional homeostasis and the malignant phenotype},
	journal = {Cancer Cell},
	volume = {8},
	number = {3},
	pages = {241-254},
	year = {2005},
	issn = {1535-6108},
	doi = {https://doi.org/10.1016/j.ccr.2005.08.010},
	author = {Matthew J. Paszek and Nastaran Zahir and Kandice R. Johnson and Johnathon N. Lakins and Gabriela I. Rozenberg and Amit Gefen and Cynthia A. Reinhart-King and Susan S. Margulies and Micah Dembo and David Boettiger and Daniel A. Hammer and Valerie M. Weaver}
}

@article{Kakkada2018,
	title = {Cell Growth Rate Dictates the Onset of Glass to Fluidlike Transition and Long Time Superdiffusion in an Evolving Cell Colony},
	author = {Malmi-Kakkada, Abdul N. and Li, Xin and Samanta, Himadri S. and Sinha, Sumit and Thirumalai, D.},
	journal = {Phys. Rev. X},
	volume = {8},
	issue = {2},
	pages = {021025},
	numpages = {21},
	year = {2018},
	month = {Apr},
	publisher = {American Physical Society},
	doi = {10.1103/PhysRevX.8.021025},
	url = {https://link.aps.org/doi/10.1103/PhysRevX.8.021025}
}

@article{Gnan2019,
	author  = {Gnan, Nicoletta and Zaccarelli, Emanuela},
	title   = {The microscopic role of deformation in the dynamics of soft colloids},
	journal = {Nature Physics},
	volume  = {15},
	pages   = {683--688},
	year    = {2019},
	doi     = {10.1038/s41567-019-0480-1}
}

@article{Kirkpatrick1988,
  title = {Comparison between dynamical theories and metastable states in regular and glassy mean-field spin models with underlying first-order-like phase transitions},
  author = {Kirkpatrick, T. R. and Thirumalai, D.},
  journal = {Phys. Rev. A},
  volume = {37},
  issue = {11},
  pages = {4439--4448},
  numpages = {0},
  year = {1988},
  month = {Jun},
  publisher = {American Physical Society},
  doi = {10.1103/PhysRevA.37.4439},
}

@article{Dempster1977,
  author    = {A. P. Dempster and N. M. Laird and D. B. Rubin},
  title     = {Maximum Likelihood from Incomplete Data via the EM Algorithm},
  journal   = {Journal of the Royal Statistical Society: Series B (Methodological)},
  volume    = {39},
  number    = {1},
  pages     = {1--38},
  year      = {1977},
  publisher = {Wiley for the Royal Statistical Society},
  issn      = {0035-9246}
}

@Article{Fuhs2022,
author={Fuhs, Thomas
and Wetzel, Franziska
and Fritsch, Anatol W.
and Li, Xinzhi
and Stange, Roland
and Pawlizak, Steve
and Kie{\ss}ling, Tobias R.
and Morawetz, Erik
and Grosser, Steffen
and Sauer, Frank
and Lippoldt, J{\"u}rgen
and Renner, Frederic
and Friebe, Sabrina
and Zink, Mareike
and Bendrat, Klaus
and Braun, J{\"u}rgen
and Oktay, Maja H.
and Condeelis, John
and Briest, Susanne
and Wolf, Benjamin
and Horn, Lars-Christian
and H{\"o}ckel, Michael
and Aktas, Bahriye
and Marchetti, M. Cristina
and Manning, M. Lisa
and Niendorf, Axel
and Bi, Dapeng
and K{\"a}s, Josef A.},
title={Rigid tumours contain soft cancer cells},
journal={Nature Physics},
year={2022},
month={Dec},
day={01},
volume={18},
number={12},
pages={1510-1519},
issn={1745-2481},
doi={10.1038/s41567-022-01755-0}
}

@article{Petridou2019,
	author = {Petridou, Nicoletta I and Heisenberg, Carl‐Philipp},
	title = {Tissue rheology in embryonic organization},
	journal = {The EMBO Journal},
	volume = {38},
	number = {20},
	pages = {e102497},
	year = {2019}
}

@article {Fridtjof2024,
	article_type = {journal},
	title = {The geometric basis of epithelial convergent extension},
	author = {Brauns, Fridtjof and Claussen, Nikolas H and Lefebvre, Matthew F and Wieschaus, Eric F and Shraiman, Boris I},
	editor = {Bagnat, Michel and Campelo, Felix},
	volume = 13,
	year = 2024,
	month = {dec},
	pub_date = {2024-12-19},
	pages = {RP95521},
	doi = {10.7554/eLife.95521},
	journal = {eLife},
	issn = {2050-084X},
	publisher = {eLife Sciences Publications, Ltd},
}

@article{Verdier2009,
	title = {Review: Rheological properties of biological materials},
	journal = {Comptes Rendus Physique},
	volume = {10},
	number = {8},
	pages = {790-811},
	year = {2009},
	note = {Complex and biofluids},
	issn = {1631-0705},
	author = {Claude Verdier and Jocelyn Etienne and Alain Duperray and Luigi Preziosi}
}

@article{Friedrich2016,
	author = {Fischer-Friedrich, Elisabeth and Toyoda, Yusuke and Cattin, Cedric J. and M{\"u}ller, Daniel J. and Hyman, Anthony A. and J{\"u}licher, Frank},
	isbn = {0006-3495},
	journal = {Biophysical Journal},
	number = {3},
	pages = {589--600},
	publisher = {Elsevier},
	title = {Rheology of the Active Cell Cortex in Mitosis},
	type = {doi: 10.1016/j.bpj.2016.06.008},
	volume = {111},
	year = {2016}}

@article{Charlotte2017,
	author = {Alibert, Charlotte and Goud, Bruno and Manneville, Jean-Baptiste},
	title = {Are cancer cells really softer than normal cells?},
	journal = {Biology of the Cell},
	volume = {109},
	number = {5},
	pages = {167-189},
	doi = {https://doi.org/10.1111/boc.201600078},
	year = {2017}
}

@article{Lekka2016,
	author = {Lekka, Ma{\l}gorzata},
	doi = {10.1007/s12668-016-0191-3},
	isbn = {2191-1649},
	journal = {BioNanoScience},
	number = {1},
	pages = {65--80},
	title = {Discrimination Between Normal and Cancerous Cells Using AFM},
	volume = {6},
	year = {2016}}

@article{Guan2021,
	title = {Unified description of compressive modulus revealing multiscale mechanics of living cells},
	author = {Guan, Dongshi and Shen, Yusheng and Zhang, Rui and Huang, Pingbo and Lai, Pik-Yin and Tong, Penger},
	journal = {Phys. Rev. Res.},
	volume = {3},
	issue = {4},
	pages = {043166},
	numpages = {13},
	year = {2021},
	month = {Dec},
	publisher = {American Physical Society},
	doi = {10.1103/PhysRevResearch.3.043166}
}

@article{Jain2014,
   author = "Jain, Rakesh K. and Martin, John D. and Stylianopoulos, Triantafyllos",
   title = "The Role of Mechanical Forces in Tumor Growth and Therapy", 
   journal= "Annual Review of Biomedical Engineering",
   year = "2014",
   volume = "16",
   number = "Volume 16, 2014",
   pages = "321-346",
   publisher = "Annual Reviews",
   issn = "1545-4274"
  }

@article{Dakhil2016,
	author = {Dakhil, Haider and Gilbert, Daniel F. and Malhotra, Deepika and Limmer, Anja and Engelhardt, Hannes and Amtmann, Anette and Hansmann, Jan and H{\"u}bner, Holger and Buchholz, Rainer and Friedrich, Oliver and Wierschem, Andreas},
	doi = {10.1007/s00397-016-0936-5},
	isbn = {1435-1528},
	journal = {Rheologica Acta},
	number = {7},
	pages = {527--536},
	title = {Measuring average rheological quantities of cell monolayers in the linear viscoelastic regime},
	volume = {55},
	year = {2016}}

@article{Lee2022,
	author = {Lee, Suhyang  and Bashir, Khawaja Muhammad Imran  and Jung, Dong Hee  and Basu, Santanu Kumar  and Seo, Gayeon  and Cho, Man-Gi  and Wierschem, Andreas },
	title = {Measuring the linear viscoelastic regime of MCF-7 cells with a monolayer rheometer in the presence of microtubule-active anti-cancer drugs at high concentrations},
	journal = {Interface Focus},
	volume = {12},
	number = {6},
	pages = {20220036},
	year = {2022},
	doi = {10.1098/rsfs.2022.0036}
}

@article{Kayal2026,
   author = "Kayal, Sayantani and Nguyen, Anh Q. and Bi, Dapeng",
   title = "The Rheology of Living Tissues: From Cells to Organismal Mechanics", 
   journal= "Annual Review of Condensed Matter Physics",
   year = "2026",
   volume = "17",
   number = "Volume 17, 2026",
   pages = "285-304",
   doi = "https://doi.org/10.1146/annurev-conmatphys-071125-054711",
   publisher = "Annual Reviews",
   issn = "1947-5462",
  }

@article{Ozawa2020,
  title = {Role of fluctuations in the yielding transition of two-dimensional glasses},
  author = {Ozawa, Misaki and Berthier, Ludovic and Biroli, Giulio and Tarjus, Gilles},
  journal = {Phys. Rev. Res.},
  volume = {2},
  issue = {2},
  pages = {023203},
  numpages = {10},
  year = {2020},
  month = {May},
  publisher = {American Physical Society},
  doi = {10.1103/PhysRevResearch.2.023203},
}

@article{Varnik2004,
    author = {Varnik, F. and Bocquet, L. and Barrat, J.L.},
    title = {A study of the static yield stress in a binary Lennard-Jones glass},
    journal = {The Journal of Chemical Physics},
    volume = {120},
    number = {6},
    pages = {2788-2801},
    year = {2004},
    month = {02},
    issn = {0021-9606},
    doi = {10.1063/1.1636451},
}

@article{Hu2023,
  author    = {X. Hu and N. Liu and V. Jambur and others},
  title     = {Amorphous shear bands in crystalline materials as drivers of plasticity},
  journal   = {Nature Materials},
  year      = {2023},
  volume    = {22},
  pages     = {1071--1077},
  doi       = {10.1038/s41563-023-01597-y},
  publisher = {Springer Nature}
}

@article{Wyss2007,
  title = {Strain-Rate Frequency Superposition: A Rheological Probe of Structural Relaxation in Soft Materials},
  author = {Wyss, Hans M. and Miyazaki, Kunimasa and Mattsson, Johan and Hu, Zhibing and Reichman, David R. and Weitz, David A.},
  journal = {Phys. Rev. Lett.},
  volume = {98},
  issue = {23},
  pages = {238303},
  numpages = {4},
  year = {2007},
  month = {Jun},
  publisher = {American Physical Society},
  doi = {10.1103/PhysRevLett.98.238303},
}

@article{Ruscher2021,
  author    = {Christoph Ruscher and J{\"o}rg Rottler},
  title     = {Avalanches in the Athermal Quasistatic Limit of Sheared Amorphous Solids: An Atomistic Perspective},
  journal   = {Tribology Letters},
  year      = {2021},
  volume    = {69},
  pages     = {64},
  doi       = {10.1007/s11249-021-01439-5},
  publisher = {Springer}
}

@article{Morse2021,
author = {Peter K. Morse  and Sudeshna Roy  and Elisabeth Agoritsas  and Ethan Stanifer  and Eric I. Corwin  and M. Lisa Manning },
title = {A direct link between active matter and sheared granular systems},
journal = {Proceedings of the National Academy of Sciences},
volume = {118},
number = {18},
pages = {e2019909118},
year = {2021},
doi = {10.1073/pnas.2019909118}}

@article{Maloney2006,
  title = {Amorphous systems in athermal, quasistatic shear},
  author = {Maloney, Craig E. and Lema\^{\i}tre, Ana\"el},
  journal = {Phys. Rev. E},
  volume = {74},
  issue = {1},
  pages = {016118},
  numpages = {22},
  year = {2006},
  month = {Jul},
  publisher = {American Physical Society},
  doi = {10.1103/PhysRevE.74.016118},
}

@article{Fiocco2015,
  author    = {Davide Fiocco and Giuseppe Foffi and Srikanth Sastry},
  title     = {Memory effects in schematic models of glasses subjected to oscillatory deformation},
  journal   = {Journal of Physics: Condensed Matter},
  year      = {2015},
  volume    = {27},
  number    = {19},
  pages     = {194130},
  doi       = {10.1088/0953-8984/27/19/194130},
  publisher = {IOP Publishing}
}

@article{Leishangthem2017,
  author    = {Prasanth Leishangthem and Anoop Parmar and Srikanth Sastry},
  title     = {The yielding transition in amorphous solids under oscillatory shear deformation},
  journal   = {Nature Communications},
  year      = {2017},
  volume    = {8},
  pages     = {14653},
  doi       = {10.1038/ncomms14653},
  publisher = {Nature Publishing Group}
}

@article{Howell1999,
  title = {Stress Fluctuations in a 2D Granular Couette Experiment: A Continuous Transition},
  author = {Howell, Daniel and Behringer, R. P. and Veje, Christian},
  journal = {Phys. Rev. Lett.},
  volume = {82},
  issue = {26},
  pages = {5241--5244},
  numpages = {0},
  year = {1999},
  month = {Jun},
  publisher = {American Physical Society},
  doi = {10.1103/PhysRevLett.82.5241},
}

@article{Kabla2007,
author = {Alexandre Kabla and Georges Debr{\'e}geas},
title = {Quasi-static rheology of foams. Part 1. Oscillating strain},
journal = {Journal of Fluid Mechanics},
year = {2007},
volume = {587},
pages = {23--44},
doi = {10.1017/S0022112007007264},
publisher = {Cambridge University Press}
}

@article{Reboucas2025,
title = {Modeling drop deformations and rheology of dilute to dense emulsions},
journal = {Current Opinion in Colloid \& Interface Science},
volume = {77},
pages = {101904},
year = {2025},
issn = {1359-0294},
doi = {https://doi.org/10.1016/j.cocis.2025.101904},
author = {Rodrigo B. Reboucas and Nadia N. Nikolova and Vivek Sharma}
}

@Article{Deptula2020,
author={Deptu{\l}a, Piotr
and {\L}ysik, Dawid
and Pogoda, Katarzyna
and Cie{\'{s}}luk, Mateusz
and Namiot, Andrzej
and Mystkowska, Joanna
and Kr{\'o}l, Grzegorz
and G{\l}uszek, Stanis{\l}aw
and Janmey, Paul A.
and Bucki, Robert},
title={Tissue Rheology as a Possible Complementary Procedure to Advance Histological Diagnosis of Colon Cancer},
journal={ACS Biomaterials Science {\&} Engineering},
year={2020},
month={Oct},
day={12},
publisher={American Chemical Society},
volume={6},
number={10},
pages={5620-5631},
doi={10.1021/acsbiomaterials.0c00975},
}

@article{Fernandez2007,
	doi = {10.1088/1367-2630/9/11/419},
	year = {2007},
	month = {nov},
	publisher = {},
	volume = {9},
	number = {11},
	pages = {419},
	author = {Fernández, Pablo and Heymann, Lutz and Ott, Albrecht and Aksel, Nuri and Pullarkat, Pramod A},
	title = {Shear rheology of a cell monolayer},
	journal = {New Journal of Physics}
}

@Article{Thibaut2024,
	author ="Divoux, Thibaut and Agoritsas, Elisabeth and Aime, Stefano and Barentin, Catherine and Barrat, Jean-Louis and Benzi, Roberto and Berthier, Ludovic and Bi, Dapeng and Biroli, Giulio and Bonn, Daniel and Bourrianne, Philippe and Bouzid, Mehdi and Del Gado, Emanuela and Delanoë-Ayari, Hélène and Farain, Kasra and Fielding, Suzanne and Fuchs, Matthias and van der Gucht, Jasper and Henkes, Silke and Jalaal, Maziyar and Joshi, Yogesh M. and Lemaître, Anaël and Leheny, Robert L. and Manneville, Sébastien and Martens, Kirsten and Poon, Wilson C. K. and Popović, Marko and Procaccia, Itamar and Ramos, Laurence and Richards, James A. and Rogers, Simon and Rossi, Saverio and Sbragaglia, Mauro and Tarjus, Gilles and Toschi, Federico and Trappe, Véronique and Vermant, Jan and Wyart, Matthieu and Zamponi, Francesco and Zare, Davoud",
	title  ="Ductile-to-brittle transition and yielding in soft amorphous materials: perspectives and open questions",
	journal  ="Soft Matter",
	year  ="2024",
	volume  ="20",
	issue  ="35",
	pages  ="6868-6888",
	publisher  ="The Royal Society of Chemistry",
	doi  ="10.1039/D3SM01740K"}

@article{Falk1998,
	title = {Dynamics of viscoplastic deformation in amorphous solids},
	author = {Falk, M. L. and Langer, J. S.},
	journal = {Phys. Rev. E},
	volume = {57},
	issue = {6},
	pages = {7192--7205},
	numpages = {0},
	year = {1998},
	month = {Jun},
	publisher = {American Physical Society},
	doi = {10.1103/PhysRevE.57.7192}
}

@article{Popovic21NJP,
  title={Inferring the flow properties of epithelial tissues from their geometry},
  author={Popovi{\'c}, Marko and Druelle, Valentin and Dye, Natalie A and J{\"u}licher, Frank and Wyart, Matthieu},
  journal={New Journal of Physics},
  volume={23},
  number={3},
  pages={033004},
  year={2021},
  publisher={IOP Publishing}
}

@article{Espina23JCS,
  title={Response of cells and tissues to shear stress},
  author={Espina, Jaime A and Cordeiro, Marilia H and Milivojevic, Milan and Paji{\'c}-Lijakovi{\'c}, Ivana and Barriga, Elias H},
  journal={Journal of Cell Science},
  volume={136},
  number={18},
  pages={jcs260985},
  year={2023},
  publisher={The Company of Biologists Ltd}
}

@article{Nguyen25NatComm,
  title={Origin of yield stress and mechanical plasticity in model biological tissues},
  author={Nguyen, Anh Q and Huang, Junxiang and Bi, Dapeng},
  journal={Nature Communications},
  volume={16},
  number={1},
  pages={3260},
  year={2025},
  publisher={Nature Publishing Group UK London}
}

@article{Lin24PNAS,
  title={Scaling description of the yielding transition in soft amorphous solids at zero temperature},
  author={Lin, Jie and Lerner, Edan and Rosso, Alberto and Wyart, Matthieu},
  journal={Proceedings of the National Academy of Sciences},
  volume={111},
  number={40},
  pages={14382--14387},
  year={2014},
  publisher={National Academy of Sciences}
}

@article{Gu2022,
	title = {Label-free cell tracking enables collective motion phenotyping in epithelial monolayers},
	journal = {iScience},
	volume = {25},
	number = {7},
	pages = {104678},
	year = {2022},
	issn = {2589-0042},
	doi = {https://doi.org/10.1016/j.isci.2022.104678},
	author = {Shuyao Gu and Rachel M. Lee and Zackery Benson and Chenyi Ling and Michele I. Vitolo and Stuart S. Martin and Joe Chalfoun and Wolfgang Losert}
}

@article{Utter2008,
	title = {Experimental Measures of Affine and Nonaffine Deformation in Granular Shear},
	author = {Utter, Brian and Behringer, R. P.},
	journal = {Phys. Rev. Lett.},
	volume = {100},
	issue = {20},
	pages = {208302},
	numpages = {4},
	year = {2008},
	month = {May},
	publisher = {American Physical Society},
}

@article{Lee2013,
year = {2013},
month = {feb},
publisher = {IOP Publishing},
volume = {15},
number = {2},
pages = {025036},
author = {Lee, Rachel M and Kelley, Douglas H and Nordstrom, Kerstin N and Ouellette, Nicholas T and Losert, Wolfgang},
title = {Quantifying stretching and rearrangement in epithelial sheet migration},
journal = {New Journal of Physics}
}

@article{Das2025,
  title        = {Controlling the Morphologies and Dynamics in Three-Dimensional Tissues},
  author       = {Das, Rajsekhar and Li, Xin and Sinha, Sumit and Thirumalai, D.},
  journal      = {arXiv, 2505.06168},
  year         = {2025},
  eprint       = {2505.06168},
  primaryClass = {cond-mat.soft}
}

@article{Herschel1926,
	author    = {William H. Herschel and Ronald Bulkley},
	title     = {Measurement of Consistency as Applied to Rubber-Benzene Solutions},
	journal   = {Proceedings of the American Society for Testing Materials},
	volume    = {26},
	number    = {2},
	pages     = {621--633},
	year      = {1926},
}

@article{Drasdo2005,
	doi = {10.1088/1478-3975/2/3/001},
	year = {2005},
	month = {jul},
	publisher = {},
	volume = {2},
	number = {3},
	pages = {133},
	author = {Drasdo, Dirk and Höhme, Stefan},
	title = {A single-cell-based model of tumor growth in vitro: monolayers and spheroids},
	journal = {Physical Biology}
}

@article{Guo1995,
	author = {Guo, Zhuyan and Thirumalai, D.},
	title = {Kinetics of protein folding: Nucleation mechanism, time scales, and pathways},
	journal = {Biopolymers},
	volume = {36},
	number = {1},
	pages = {83-102},
	doi = {https://doi.org/10.1002/bip.360360108},
	year = {1995}
}

@article{Lees1972,
	doi = {10.1088/0022-3719/5/15/006},
	year = {1972},
	month = {aug},
	publisher = {},
	volume = {5},
	number = {15},
	pages = {1921},
	author = {A W Lees and S F Edwards},
	title = {The computer study of transport processes under extreme conditions},
	journal = {Journal of Physics C: Solid State Physics}
}

@article {Das2024,
	article_type = {journal},
	title = {Free volume theory explains the unusual behavior of viscosity in a non-confluent tissue during morphogenesis},
	author = {Das, Rajsekhar and Sinha, Sumit and Li, Xin and Kirkpatrick, TR and Thirumalai, D},
	editor = {Kruse, Karsten and Walczak, Aleksandra M},
	volume = 12,
	year = 2024,
	month = {jan},
	pub_date = {2024-01-19},
	pages = {RP87966},
	citation = {eLife 2024;12:RP87966},
	doi = {10.7554/eLife.87966},
	journal = {eLife},
	issn = {2050-084X},
	publisher = {eLife Sciences Publications, Ltd},
}

@article{Sharma25NatPhys,
  title={Activity-induced annealing leads to a ductile-to-brittle transition in amorphous solids},
  author={Sharma, Rishabh and Karmakar, Smarajit},
  journal={Nature Physics},
  volume={21},
  number={2},
  pages={253--261},
  year={2025},
  publisher={Nature Publishing Group UK London}
}

@article{Berthier25NatRevPhys,
  title={Yielding and plasticity in amorphous solids},
  author={Berthier, Ludovic and Biroli, Giulio and Manning, Lisa and Zamponi, Francesco},
  journal={Nature Reviews Physics},
  volume={7},
  number={6},
  pages={313--330},
  year={2025},
  publisher={Nature Publishing Group UK London}
}

@article{Pathmanathan2009,
	doi = {10.1088/1478-3975/6/3/036001},
	year = {2009},
	month = {apr},
	publisher = {},
	volume = {6},
	number = {3},
	pages = {036001},
	author = {Pathmanathan, P and Cooper, J and Fletcher, A and Mirams, G and Murray, P and Osborne, J and Pitt-Francis, J and Walter, A and Chapman, S J},
	title = {A computational study of discrete mechanical tissue models},
	journal = {Physical Biology}
}

@article{Schaller2005,
	title = {Multicellular tumor spheroid in an off-lattice Voronoi-Delaunay cell model},
	author = {Schaller, Gernot and Meyer-Hermann, Michael},
	journal = {Phys. Rev. E},
	volume = {71},
	issue = {5},
	pages = {051910},
	numpages = {16},
	year = {2005},
	month = {May},
	publisher = {American Physical Society},
	doi = {10.1103/PhysRevE.71.051910},
}

@article{Ryoichi1998,
	title = {Dynamics of highly supercooled liquids: Heterogeneity, rheology, and diffusion},
	author = {Yamamoto, Ryoichi and Onuki, Akira},
	journal = {Phys. Rev. E},
	volume = {58},
	issue = {3},
	pages = {3515--3529},
	numpages = {0},
	year = {1998},
	month = {Sep},
	publisher = {American Physical Society},
	doi = {10.1103/PhysRevE.58.3515}
}

@article{Tang2023,
	title = {The brittle-to-ductile transition in aluminosilicate glasses is driven by topological and dynamical heterogeneity},
	journal = {Acta Materialia},
	volume = {247},
	pages = {118740},
	year = {2023},
	issn = {1359-6454},
	doi = {https://doi.org/10.1016/j.actamat.2023.118740},
	author = {Longwen Tang and Morten M. Smedskjaer and Mathieu Bauchy}
}

@article{Shrivastav2020,
	author    = {Gaurav P. Shrivastav and Gerhard Kahl},
	title     = {On the stress overshoot in cluster crystals under shear},
	journal   = {Condensed Matter Physics},
	volume    = {23},
	number    = {2},
	pages     = {23801},
	year      = {2020},
	doi       = {10.5488/CMP.23.23801}
}

@book{Gumbel2013,
	author    = {E. J. Gumbel},
	title     = {Statistics of Extremes},
	publisher = {Echo Point Books \& Media},
	year      = {2013},
	isbn      = {978-1-62654-987-6},
	pages     = {396}
}

@Article{McCord2025,
author ="McCord, Molly and Notbohm, Jacob",
title  ="Measurement of tissue viscosity to relate force and motion in collective cell migration",
journal  ="Soft Matter",
year  ="2026",
volume  ="22",
issue  ="2",
pages  ="458-473",
publisher  ="The Royal Society of Chemistry",
doi  ="10.1039/D5SM01139F"}
\newpage
\clearpage
\thispagestyle{empty}
\vspace*{\fill}
\begin{center}
    \includegraphics[
        width=0.92\textwidth,
        height=0.95\textheight,
        keepaspectratio
    ]{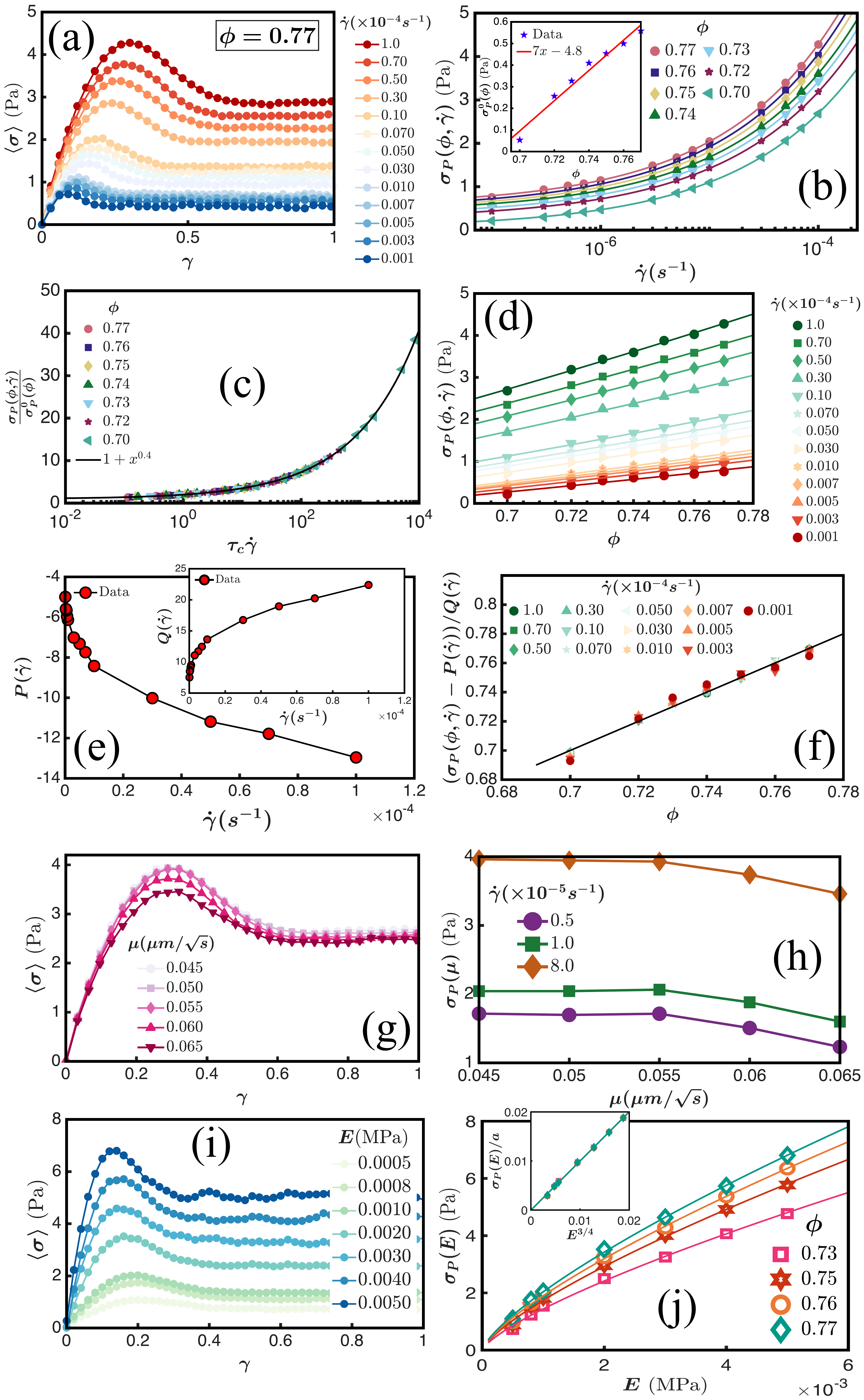}
\end{center}
\vspace*{\fill}
\clearpage
\vspace*{\fill}
\begin{center}
\begin{minipage}{\textwidth}
    \captionof{figure}{ {\bf Yielding transition in 3D polydisperse system of cells}: (a) Ensemble-averaged stress-strain profiles for volume fraction $\phi = 0.77$ and $\mu = 0.045\mu m/\sqrt{s}$ for decreasing value of $\dot{\gamma}$ from top to bottom). The yield stress, $\sigma_P(\phi,\dot{\gamma})$, identified as the peak in the curves, occurs before reaching the steady state. (b) $\sigma_P(\phi,\dot{\gamma})$ versus $\dot{\gamma}$ for various $\phi$. Symbols are the simulation data and solid lines are power law fits to Eq.~(\ref{HBEq}) with $n = 0.4$. Inset: Dependence of the yield stress $\sigma_P^0(\phi)$ on $\phi$. Points are from simulations, and the red line is the linear fit. (c) Collapse of the data in Fig.~(\ref{YieldingTrans}b). The solid black line is the plot of the function $f(x) = 1 + x^{0.4}$ (Eq.~(\ref{HBEqCollapse})). (d) $\sigma_P(\phi,\dot{\gamma})$  as a function of $\phi$ for a fixed $\dot{\gamma}$ shows a linear variation. Symbols are simulation data, and the solid lines are the fit to Eq.~(\ref{HBEqvariationPhi}).  (e) $P(\dot{\gamma})$ as function of $\dot{\gamma}$. Inset: Variation of $Q(\dot{\gamma})$ with  $\dot{\gamma}$ (Eq.~(\ref{HBEqvariationPhi})). (f) Scaling collapse of the data in Fig.~(\ref{YieldingTrans}d). The line is the plot of the function $f(x) = x$. The other simulation parameters for Fig.(a-f) are: $E = 0.001MPa, \mu = 0.045\mu m/\sqrt{s}, \Sigma = 8.5\%$. (g) Ensemble-averaged stress versus strain for five different $\mu$ values for $\dot{\gamma} = 8 \times 10^{-5} s^{-1}$. (h) The yield stress, $\sigma_P(\mu)$ versus $\mu$. $\sigma_P(\mu)$ increases as $\mu$ decreases, and finally saturates. (i) Ensemble-averaged stress as a function of strain at multiple $E$ values. An increase in $E$ leads to an increase in the yield stress in the stress overshoot curve. (j) $\sigma_{P}(E)$ as a function of $E$. Points are the simulation data.  Lines are the fits to a power law, $f(x) = a x^{3/4}$. Inset: Plot of $\sigma_{P}(E)/a$ versus $E^{3/4}$, results in the curves in Fig.~(\ref{YieldingTrans}j) collapsing onto a single master curve. }
    \label{YieldingTrans}
\end{minipage}
\end{center}
\vspace*{\fill}
\clearpage
\thispagestyle{empty}
\vspace*{\fill}
\begin{center}
    \includegraphics[
        width=\textwidth,
        height=\textheight,
        keepaspectratio
    ]{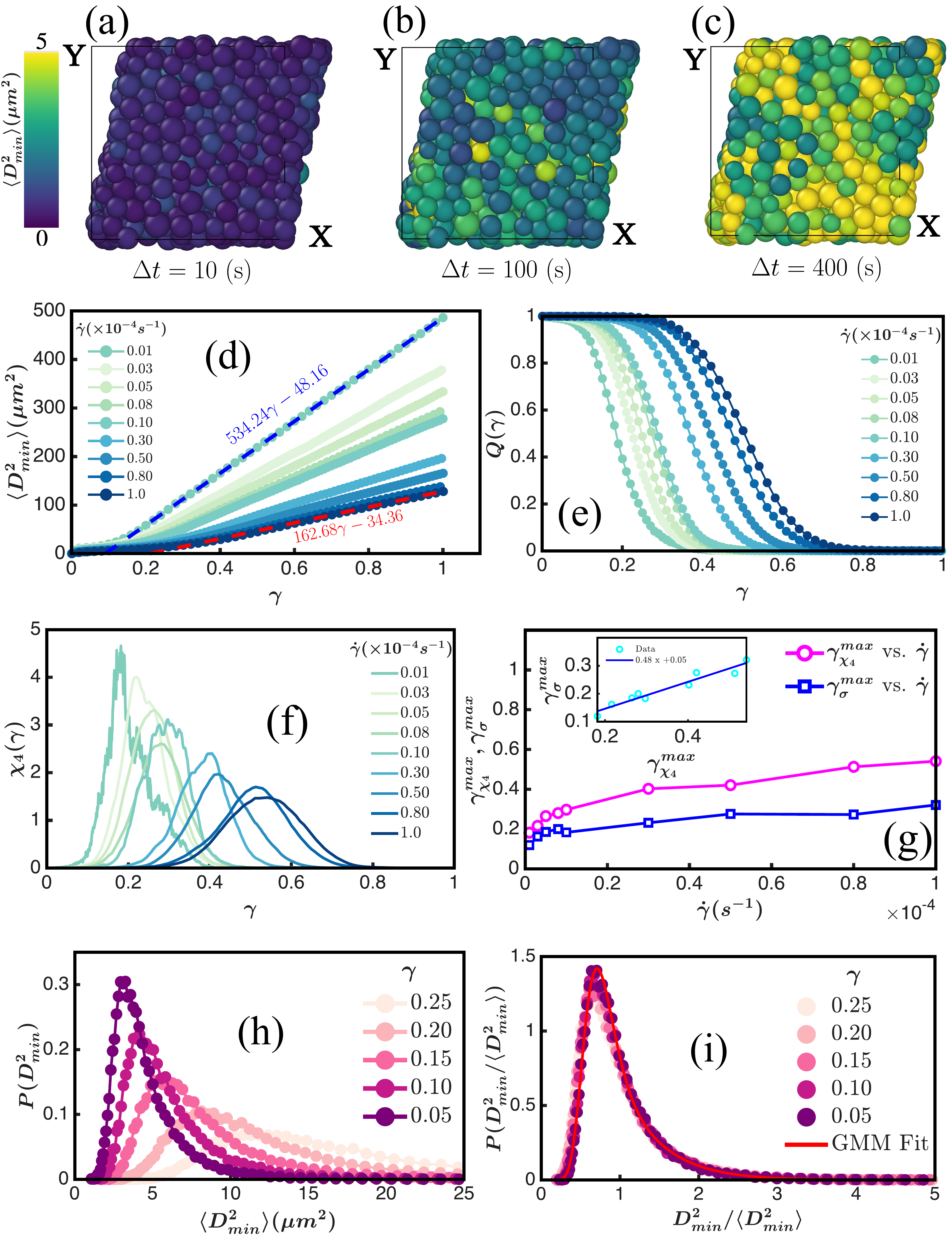}
\end{center}
\vspace*{\fill}
\clearpage
\vspace*{\fill}
\begin{center}
\begin{minipage}{\textwidth}
    \captionof{figure}{{\bf Quantifying the non-affine displacements}: (a-c) Configurations of the cells in the $xy$-plane. Cells are colored by their corresponding non-affine displacement, $D^2_{min}(t, \Delta t)$, at a strain, $\gamma = 0.2$, for three different values of $\Delta t$. As we increase $\Delta t$ (from left to right), the rearrangements increase. The color bar on the left shows the scale of $D^2_{min}(t, \Delta t)$. (d) Average $D^2_{min}(0, \Delta t)$ for $\dot{\gamma}$ varying from $10^{-4} - 10^{-6}s^{-1}$. Decreasing $\dot{\gamma}$ increases the average value because non-affine displacements are larger at low shear rates. These curves vary linearly with $\gamma$ after the intermediate plateau. We have shown the linear fits of the highest and the lowest shear rate data by red and blue lines, respectively. (e) Self-overlap function, $Q(\gamma)$ defined from the average $D^2_{min}(0, \Delta t)$ for a cut-off, $D_{c}$, chosen at around the plateau value.  $Q(\gamma)$ decays faster for smaller $\dot{\gamma}$ and slower for larger $\dot{\gamma}$. (f) Four-point susceptibility, $\chi_4(\gamma)$, defined using the fluctuations in $Q(\gamma)$. Fluctuations $\chi_4(\gamma)$ increases as $\dot{\gamma}$ decreases. (g) Positions of maximum of $\chi_4(\gamma)$ and the position of maximum of stress, $\sigma$, versus $\dot{\gamma}$. They tend to saturate at large $\dot{\gamma}$. Inset: Positions of maximum of $\chi_4(\gamma)$ and the position of maximum of stress, $\sigma$, vary linearly with each other. The linear fit is shown by the blue line in the inset. (h) $P(D^2_{\min})$ versus $D^2_{\min}$ measured at various strains with respect to the unstrained reference state at $t = 0$. The peak height of the distribution decreases as $\gamma$ increases. (i) The curves collapse into a single master curve upon scaling $D^2_{min}$ by $\langle D^2_{min} \rangle$.  The fit to the Gaussian mixture model (GMM)  (Eq.~(\ref{GMM})) is shown by the solid red line. $D^2_{min}$ is measured with respect to the unstrained configuration. The other parameters are: $\phi = 0.77$, $E = 0.001$ MPa, $\mu = 0.045\mu m/\sqrt{s}$, $\Sigma = 8.5\%$.}
    \label{D2minProps}
\end{minipage}
\end{center}
\vspace*{\fill}
\clearpage
\thispagestyle{empty}
\vspace*{\fill}
\begin{center}
    \includegraphics[
        width=\textwidth,
        height=\textheight,
        keepaspectratio
    ]{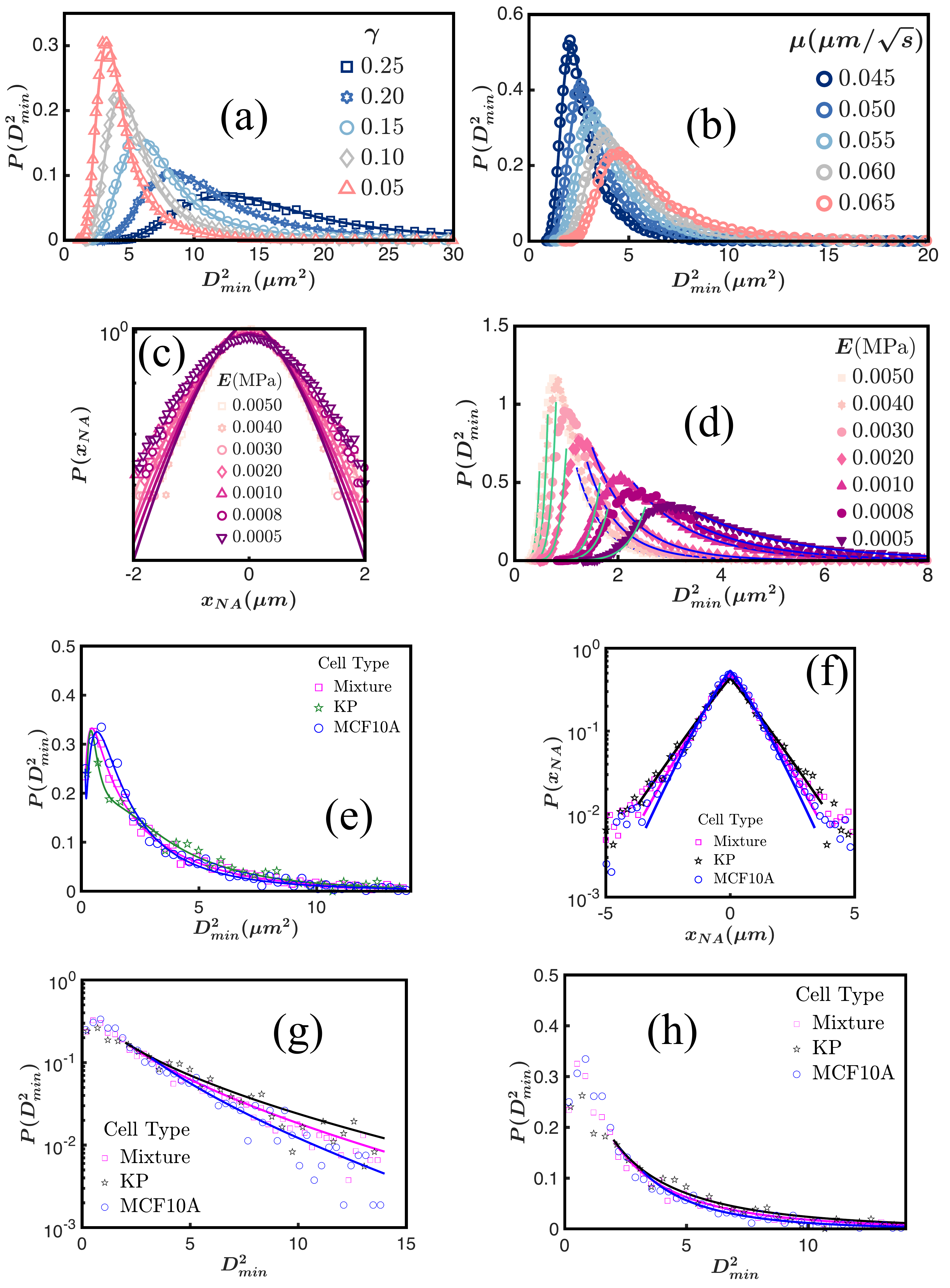}
\end{center}
\vspace*{\fill}
\clearpage
\vspace*{\fill}
\begin{center}
\begin{minipage}{\textwidth}
    \captionof{figure}{{\bf Theoretical forms of \bm{$P(D^2_{min})$} versus \bm{$D^2_{\min}$} for constant activity and stiffness model:} (a) $P(D^2_{\min})$ versus $D^2_{\min}$ measured at five different values of $\gamma$. The points are from the simulations, and the corresponding lines are the fit with the GMM model with $K = 2$. The rest  of the parameters used here are: $\phi = 0.77, \dot{\gamma} = 10^{-5}s^{-1}, \mu = 0.045\mu m/\sqrt{s}, \Sigma = 8.5\%$, and $E = 0.001$ MPa. (b) $P(D^2_{\min})$ versus $D^2_{\min}$ for five different values of $\mu$ measured at $\gamma = 0.2$ with $\Delta t = 100s$.  The points are from the simulations, and the corresponding lines are the fit with the GMM model. The rest  of the parameters used here are: $\phi = 0.77, \dot{\gamma} = 10^{-5}s^{-1}, \Sigma = 8.5\%$, and $E = 0.001 MPa$. As shown in the plots (a-b), the fit of the simulation data of $P(D^2_{\min})$ versus $D^2_{\min}$ with GMM is excellent. (c) Probability distribution of the X-component of non-affine displacements, $P(x_{NA})$, versus $x_{NA}$ on a semi-log scale for a set of $E$ values.  The points are from simulations, and the lines are fits with a non-Gaussian form with exponent $\alpha$. We found that $\alpha$ varies between $\sim 1.46-1.86$. (d) $P(D^2_{\min})$ versus $D^2_{\min}$ obtained for the same set of $E$ at $\gamma = 0.2$ with $\Delta t = 100s$.  As $E$ increases, peak height increases. We fit the two extreme limits of the distribution using Eq.~(\ref{D2minTheo}) with the same $\alpha$ estimated from (c). We have shown the fits for $D^2_{\min} \rightarrow \infty$ with blue lines and $D^2_{\min} \rightarrow 0$ with green lines. As shown here, our estimated limiting forms captured the growth and the decay of $P(D^2_{\min})$ nicely. For (c) and (d), the rest of the parameters are: $\phi = 0.77$, $\dot{\gamma} = 10^{-5}s^{-1}$, $\mu = 0.045\mu m/\sqrt{s}$, $\Sigma = 8.5\%$. {(e) $P(D^2_{\min})$ versus $D^2_{\min}$ measured for three different cell-types in the experiments. Symbols are the experimental data, and the corresponding lines are the fit with the GMM model. {(f)  Probability distribution of the X-component of non-affine displacements, $P(x_{NA})$ versus $x_{NA}$ on a semi-log scale for three different cell-culture conditions in experiments. The points are from experiments, and the lines are fits with a non-Gaussian form with exponent $\alpha$. The $\alpha$ values vary between $\sim 1.05-1.15$. (g) $P(D^2_{\min})$ versus $D^2_{\min}$ on a semi-log scale for three different cell-culture conditions in the experiments. We fit the large-$D^2_{min}$ limit (shown by the lines) of the distribution using Eq.~(\ref{D2minTheo}) with the $\alpha$ estimated from (f). (h) Same plot as (g) on a normal scale.} }}
    \label{FunctionalForms}
\end{minipage}
\end{center}
\vspace*{\fill}
\clearpage
\thispagestyle{empty}

\vspace*{\fill}
\begin{center}
    \includegraphics[
        width=\textwidth,
        height=\textheight,
        keepaspectratio
    ]{TestSphereSimRandomNp.pdf}
\end{center}
\vspace*{\fill}
\clearpage
\vspace*{\fill}
\begin{center}
\begin{minipage}{\textwidth}
    \captionof{figure}{\textbf{Mechanical response of the cellular system due to inclusion of randomly chosen $N_p$ cells with low stiffness and high motility: } (a) Configuration of the cells with $N_p$ randomly chosen cells (yellow) with $\mu = 0.07\mu m/\sqrt{s}$ and $E = 0.5E_0$. The rest of the cells (violet) have $\mu = 0.045\mu m/\sqrt{s}$ and $E = 5E_0$. (b) Ensemble-averaged stress versus strain for six different values of $N_p$. As the number of these soft active cells increases, the yield stress decreases. (c) The bar plot of the average $D^2_{min}$ values of MCF10A mutant and MDA-MB-231 cells. This plot is reproduced from Fig.(2C) and (2E) of Ref.\cite{Gu2022}. Right y-axis: The average $D^2_{min}$ as a function of $N_p$ is plotted with the line points. (d) $P(D^2_{min})$ versus $D^2_{min}$ for various $N_p$. As $N_p$ increases, the variability in $D^2_{min}$ increases that broadens the peak of $P(D^2_{min})$. (e) The 3D phase space diagram of $\mu_{N_p}-\mu_{rest}-\sigma_P$. (f) The corresponding 2D representation of the plot in (e). {The color bars in (e-f) represent the value of the yield stress, $\sigma_P$.}  (g) The 3D phase space diagram of $\mu_{N_p}-\mu_{rest}-\text{Maxval of }P(D^2_{min})$. (h) The corresponding 2D representation of the plot in (g). {The color bars in (g-h) represent the $\text{Maxval of }P(D^2_{min})$}. (i-l) $P(D^2_{min})$ versus $D^2_{min}$ for all cells, $N_p (= 250)$ cells, and the rest of the cells for four different sets of $\mu_{N_p}$ and $\mu_{rest}$. The parameters used here are: $\phi = 0.77, \dot{\gamma} = 10^{-5}s^{-1}, \Sigma = 8.5\%$. $D^2_{min}$ is measured at $\gamma = 0.2$ with $\Delta t = 100s$. {(m) $P(D^2_{min})$ versus $D^2_{min}$ for mixture of cells, cancerous KP cells, and non-cancerous MCF10A cells. The $P(D^2_{min})$ shows a similar trend as our simulation results for varying $\mu$ and differences in $E$ between $N_p$ cells and the rest of the cells. }}
    \label{TestSphereSimRandomNp}
\end{minipage}
\end{center}
\vspace*{\fill}
\clearpage
\thispagestyle{empty}

\vspace*{\fill}
\begin{center}
    \includegraphics[
        width=\textwidth,
        height=\textheight,
        keepaspectratio
    ]{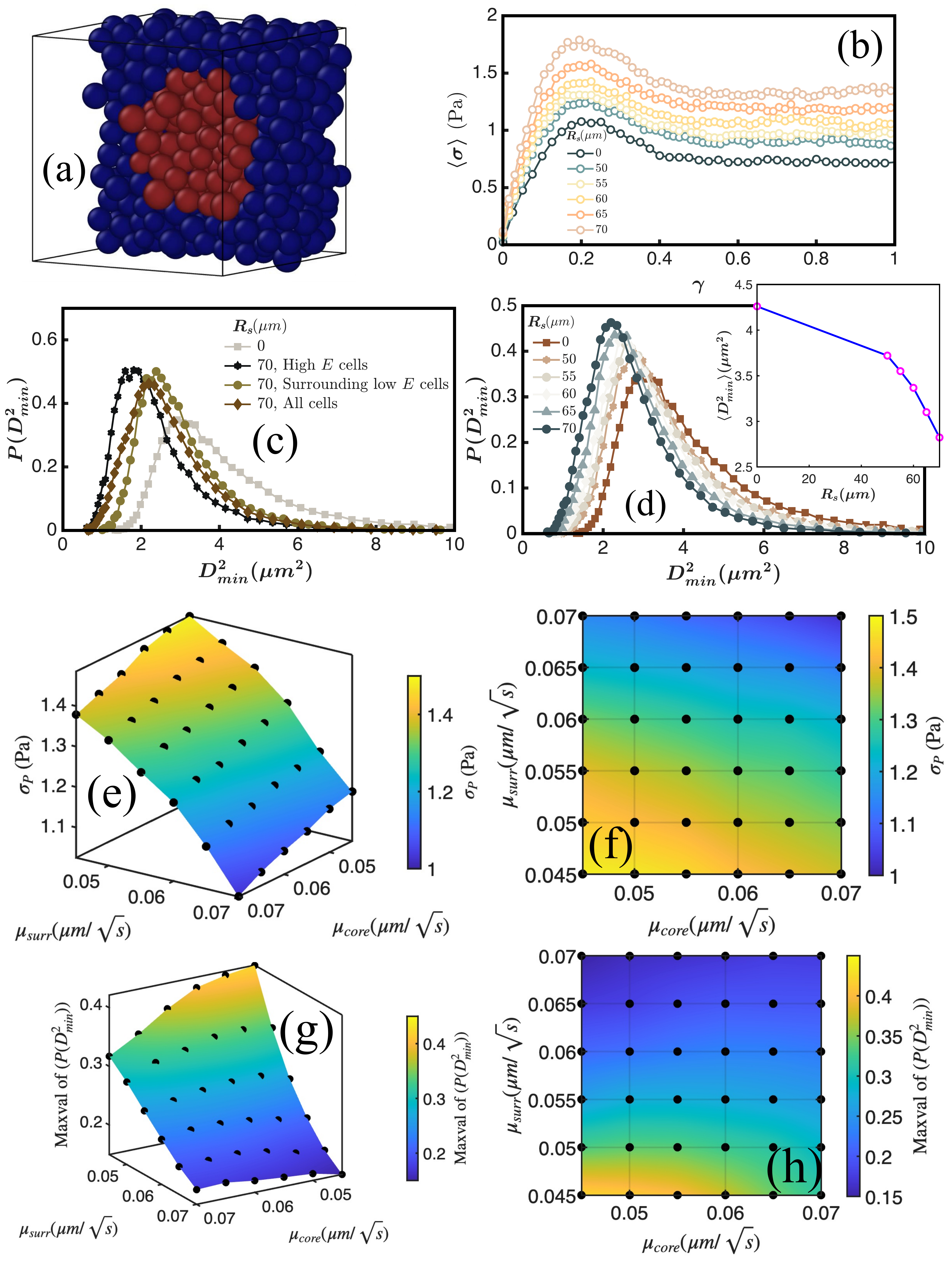}
\end{center}
\vspace*{\fill}
\clearpage
\vspace*{\fill}
\begin{center}
\begin{minipage}{\textwidth}
    \captionof{figure}{\textbf{Changes in mechanical properties upon inclusion of a spherical region of stiff cells in a sea of normal cells:} (a) A cross-section of the configuration of cells with high $E$ (red) surrounded by low $E$ (blue). Here we have simulated a spherical stiff tumor of radius $R_s$ inside a soft cellular environment. (b) Ensemble-averaged stress versus strain for six different values of $R_s$.  As $R_s$ increases, the yield stress increases as more cells become stiffer. (c) The distribution of $D^2_{min}$ for a fixed $R_s = 70\mu m$ for the stiff cells, surrounding soft cells, and all cells. $P(D^2_{min})$ for $R_s = 0$ is also shown for the reference. (d)  $P(D^2_{min})$ versus $D^2_{min}$ for various $R_s$. As $R_s$ increases, $P(D^2_{min})$ becomes sharper as the number of stiff cells increases. Inset: The average $D^2_{min}$ as a function of $R_s$.  $R_s = 0$ corresponds to the single activity and single stiffness case.  (e) The 3D phase space diagram of $\mu_{core}-\mu_{surr}-\sigma_P$. (f) The corresponding 2D representation of the plot in (e). {The colorbars in (e-f) represent the value of the yield stress, $\sigma_P$.} (g) The 3D phase space diagram of $\mu_{core}-\mu_{surr}-\text{Maxval of }P(D^2_{min})$. (h) The corresponding 2D representation of the plot in (g). {The colorbars in (g-h) show the $\text{Maxval of }P(D^2_{min})$.} The parameters used for are: $\phi = 0.77, \dot{\gamma} = 10^{-5}s^{-1}, \Sigma = 8.5\%$. $D^2_{min}$ is measured at $\gamma = 0.2$ with $\Delta t = 100s$. }
    \label{TestSphereSimSphRadius}
\end{minipage}
\end{center}
\vspace*{\fill}
\clearpage
\thispagestyle{empty}
\vspace*{\fill}
\begin{center}
    \includegraphics[
        width=\textwidth,
        height=\textheight,
        keepaspectratio
    ]{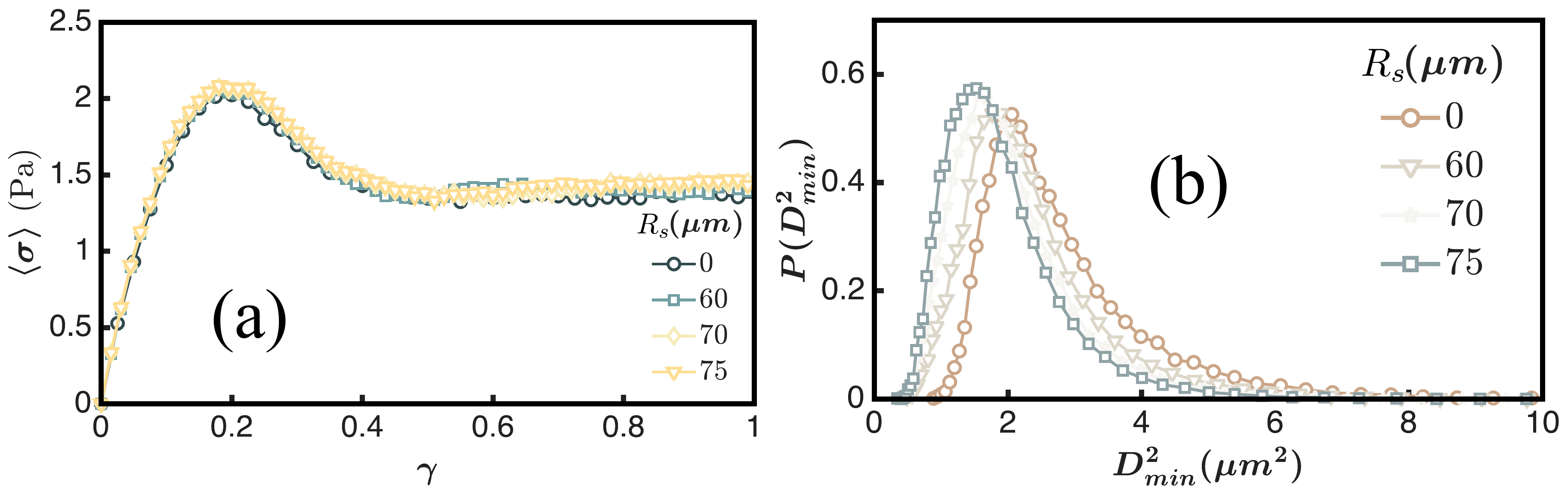}
\end{center}
\vspace*{\fill}
\clearpage
\vspace*{\fill}
\begin{center}
\begin{minipage}{\textwidth}
    \captionof{figure}{\textbf{Influence of a spherical region of slow cells:} (a)  Ensemble-averaged stress versus strain for four different values of $R_s$. Now, in this simulation study, we made the activity, $\mu$, of the cells smaller $(\mu = 0.045\mu m/\sqrt{s})$ than that of the surrounding cells $(\mu = 0.065\mu m/\sqrt{s})$ while keeping $E$ constant for all the cells. The yield stress increases a little with $R_s$. (b) $P(D^2_{min})$ versus $D^2_{min}$ for various $R_s$. The peak of $P(D^2_{min})$ increases with $R_s$. The parameters used here are: $\phi = 0.77, \dot{\gamma} = 10^{-5}s^{-1}, \Sigma = 8.5\%, E = 0.001 MPa$. $D^2_{min}$ is measured at $\gamma = 0.2$ with $\Delta t = 100s$. }
    \label{TestSphereSimChangingMu}
\end{minipage}
\end{center}
\vspace*{\fill}
\clearpage

\newpage
\appendix
\renewcommand{\thesection}{\Alph{section}}
\renewcommand{\theequation}{\thesection.\arabic{equation}}
\renewcommand{\thefigure}{\thesection.\arabic{figure}}
\setcounter{figure}{0}
\renewcommand{\thefigure}{A\arabic{figure}}
\makeatletter
\@addtoreset{figure}{section}
\@addtoreset{equation}{section}
\makeatother

\section{System size effects}
	\label{sec:SYSSIZE}
	We verified that the simulation results do not change significantly for three different system sizes: $N = 500, 1000$, and $5000$ (Fig.~(\ref{syssize})). The dependence of $\langle \sigma \rangle$ on $\gamma$ for $\phi = 0.76$ is shown in the top row of Fig.~(\ref{syssize}a). Fig.~(\ref{syssize}b) shows that the yield stress as a function of $\dot{\gamma}$ and the curves are well fitted by Eq.~(\ref{HBEq}) in the main text. The curves in Fig.~(\ref{syssize}b) collapse by suitable scaling, as shown in Fig.~(\ref{syssize}c), and are well fit using Eq.~(\ref{HBEqCollapse}) in the main text. We repeated the analysis for $N = 1000$ and $N = 5000$, shown in the middle (d, e, f) and bottom row (g, h, i), respectively. There is no qualitative system-size dependence in the results, except for a modest change in the values of the fitting parameter. All the curves fall on a master plot with the exponent $0.4$ (Fig.~(\ref{syssize}c, f, i)). 
	\section{Glass-like dynamics in the absence of shear}
	\label{sec:WithourShear}
	We initially prepared the system in a glassy state to obtain insights into plastic events, yielding, and non-affine displacements upon application of shear. To confirm the initial state (in the absence of shear) is glass-like, we calculated the mean-square displacement (MSD), $\Delta r^2(t)$, and the self-intermediate scattering function, $F_s(k,t)$, for five values of $\phi$ (Fig.~(\ref{woshearDynaprop}a-b)). The plateau in MSD and the slow decay in $F_s(k,t)$ suggest that the configurations are in a glassy-like state and are expected to resist shear.
	Next, we studied the effect of increasing the activity, $\mu$, of the cells. The plateau in the MSDs vanishes, and $F_s(k,t)$ decays faster as $\mu$ increases (Fig.~(\ref{woshearDynaprop}c-d)). 
	\section{Effect of polydispersity (\texorpdfstring{$\Sigma$}{Sigma}) and waiting time \texorpdfstring{$\Delta t$}{Delta t} }
	\label{sec:EffectofPoly}
	Variations in the polydispersity, $\Sigma$, do not show a significant change in the yield stress and in $P(D^2_{\min})$ (Fig.~(\ref{Polydiseffect}a) and Fig.~(\ref{Polydiseffect}b)). We have also checked the dependence of $\Delta t$ (the time spacing between two configurations) on $P(D^2_{min})$. Fig.~(\ref{Polydiseffect}c) shows that as $\Delta t$ increases, the peak of the $P(D^2_{min})$ decreases. 
	\section{Asymptotic forms of \texorpdfstring{$P(D^2_{min})$}{PD2min}
    using extreme value statistics}
	\label{sec:MFtheo}
	Utter and Behringer (UB) \cite{Utter2008} have given an approximate form of the probability distribution functions (PDFs) of $D^2_{min}$ and used it to analyze shear effects in 2D granular Couette flow.  To obtain tractable results, UB start by defining $D^2_{min}$ by summing over the individual non-affine displacements using $D^2_{min} = \frac{1}{N_g}\sum_{i=1}^{N_g} \delta r_i^2$, where $N_g$ is the number of nearest neighbor cells that contribute to $D^2_{min}$. The distribution of the individual non-affine displacement is taken to be $P(\delta r_i) \simeq A \exp(-B|\delta r_i|^\alpha)$ with $\alpha \leq 2$, where $A$ and $B$ are constants. It was assumed that the non-affine displacement distributions for individual particles are uncorrelated Gaussians ( $\alpha = 2$). In our derivation, we vary $\alpha$ to obtain a simple expression for the limiting behavior of $P(D^2_{min})$. 
	
	\noindent
	{\it Limit of $D^2_{min} \rightarrow 0$:} For $D^2_{min} \rightarrow 0$, all $\delta r_i$ must be small. In general, the full normalized $P_1(\delta r)$ should follow:
	\begin{equation}
		\label{singlePart}
		P_1(\delta r) = \frac{\alpha B}{2\Gamma(1/\alpha)} \exp(-B|\delta r_i|^\alpha),
	\end{equation}
	where $P_1(0) = \frac{\alpha B}{2\Gamma(1/\alpha)} $ is the normalization factor. The probability distribution function (PDF) for $D^2_{min}$, defined for the neighbors with $N_g$ particles, is given by \cite{Utter2008},
	\begin{equation}
		\label{PD2minStep0}
		P(D^2_{min}) = \int P_{N_g}(\delta r_1, \delta r_2 \ldots \delta r_{N_g}) \delta(D^2_{min} - \frac{1}{N_g}\sum_{i=1}^{N_g} \delta r_i^2) d\delta r_1 d\delta r_2\ldots. d\delta r_{N_g}.
	\end{equation}
	Because $ P_{N_g} (\delta r_1, \delta r_2 \ldots \delta r_{N_g})$ is taken to be uncorrelated, we can write
	\begin{equation}
		\label{PD2minStep1}
		P(D^2_{min}) = \int \Pi_{i=1}^{N_g}  P_i(\delta r_i)   \delta(D^2_{min} - \frac{1}{N_g}\sum_{i=1}^{N_g} \delta r_i^2) d\delta r_1 d\delta r_2\ldots. d\delta r_{N_g}.
	\end{equation}
	For small $\delta r$, $P_1(\delta r)$ is almost flat, and $P_1(\delta r) \simeq P_1(0)$, for any finite $\alpha$.  Hence, the product term is a constant and comes out of the integral as $P_1(0)^{N_g}$. The modified equation becomes,
	\begin{equation}
		\label{PD2minStep2}
		P(D^2_{min}) \propto  P_1(0)^{N_g}\int \delta(D^2_{min} - \frac{1}{N_g}\sum_{i=1}^{N_g} \delta r_i^2) d\delta r_1 d\delta r_2\ldots. d\delta r_{N_g}.
	\end{equation}
    The right-hand side of the integral is the area of an $ N_g$-dimensional hypersphere with radius $\tilde{R} = \sqrt{N_gD^2_{min}}$. Using the standard formula, the area is ${(N_gD^2_{min})}^{({N_g-1})/2}$. We have to account for the Jacobian factor because, $P(D^2_{min}) d(D^2_{min}) =P(\tilde{R}) d\tilde{R}$ with $P(\tilde{R}) = {(N_gD^2_{min})}^{({N_g-1})/2}$ and $dR  = \frac{\sqrt{N_g}d(D^2_{min})}{2\sqrt{D^2_{min}}}$. The final form of $P(D^2_{min})$ when $D^2_{min} \rightarrow 0$ is,
	\begin{equation}
		\label{PD2minFinalIntial}
		P(D^2_{min}) \propto P_1(0)^{N_g} ({D^2_{min}})^{N_g/2 -1}.
	\end{equation}
	
	\noindent
	{\it Large $D^2_{min}$ limit:} When $D^2_{min}$ is large, only a single non-affine displacement governs the full distribution and carries most of the weight. Let $\delta r_{N_g}$ have the largest non-affine displacement, then $\delta r_{N_g} \approx \frac{1}{N_g}\sqrt{D^2_{min}}$ while $\sum_{i=1}^{N_g-1} \delta r_i^2 \simeq 0$. Then the delta function in Eq.(~\ref{PD2minStep0}) reduces to a single integral, 
	\begin{equation}
		\label{Step1}
		\Pi_{i=1}^{N_g}  P_i(\delta r_i)  = P_{N_g}(\delta r_{N_g}) 	\Pi_{i=1}^{N_g-1}  P_i(\delta r_i \simeq 0) \propto  P_{N_g}(\delta r_{N_g}) P_1(0)^{N_g -1}.
	\end{equation}
	Since we already know the distribution of $P_{N_g}(\delta r_{N_g})$ from Eq.(~\ref{singlePart}), we can use a Jacobian transformation to get the distribution of $P(D^2_{min})$. $P(D^2_{min}) \propto P_1(0)^{N_g -1} \frac{P_{N_g}(\sqrt{D^2_{min}})}{2\sqrt{D^2_{min}}}$, the denominator comes from the transformation of variables. The final form of the tail  of the distribution is,
	\begin{equation}
		\label{PD2minFinalTail1}
		P(D^2_{min}) \propto P_1(0)^{N_g-1} \frac{\alpha B}{2\Gamma(1/\alpha)} \exp(-B({\sqrt{D^2_{min}}})^\alpha)/2\sqrt{D^2_{min}}.
	\end{equation}
	After a bit of rearrangement, we obtain
	\begin{equation}
		\label{PD2minFinalTail}
		P(D^2_{min}) \propto C \exp[-D(D^2_{min})^{\alpha/2}].
	\end{equation}
	In the above equation, there is a slowly varying $(D^2_{min})^{-1/2}$ term compared to the stretched exponential decay.

	To compare the theoretical formula to simulations, we first fit the distribution of the $X$-component of the non-affine displacement using
    \begin{equation}
		\label{PxNAEq}
		P(x_{NA}) = A_1 \exp(-B_1|x_{NA}|^\alpha),
	\end{equation}
    to estimate $\alpha$. $A_1$ and $B_1$ are two constants. Then we use the same $\alpha$, for fitting the $D^2_{min} \rightarrow \infty$ form. For our data, we took $N_g$ as a constant in {Eq.~(\ref{PD2minFinalIntial}) and in} Eq.~(\ref{D2minTheo}) (in the main text) because it does not vary much. Therefore, we have one fitting parameter, $C_1$, for the small $D^2_{min}$ part and two fitting parameters, $C_2$ and $D$, for the large $D^2_{min}$ part. In Fig.~(\ref{D2minDist}), we have computed the non-affine displacement distribution and the distribution of the $D^2_{min}$ for changing various parameters for $\Delta t = 100s$ at $\gamma = 0.2$ (close to the yielding point). First, we measured the $P(x_{NA})$ for a set of $\phi$ as shown in Fig.~(\ref{D2minDist}a). We found that $\alpha$ varies a little between $\sim 1.53-1.56$. Points are the simulation data, and the lines are the fit with the non-Gaussian form. We also obtained the distribution of $D^2_{min}$ as shown in Fig.~(\ref{D2minDist}b) for a set of $\phi$. Because decreasing $\phi$ makes the system liquid-like, the cell rearranges more frequently via non-affine displacements. Therefore, this causes a greater variability in the $D^2_{min}$ values, resulting in a wide distribution in $P(D^2_{min})$ as $\phi$ decreases. We have shown the fit for the small $D^2_{min}$  part with red lines and the large $D^2_{min}$  part with blue lines in Fig.~(\ref{D2minDist}b) using Eq.~(\ref{D2minTheo}).  
    
    Next, we varied the activity, $\mu$, of the cells and obtained $P(x_{NA})$ as shown in Fig.~(\ref{D2minDist}c), which gives rise to $\alpha \sim 1.86-1.39$. $P(D^2_{min})$ for five different $\mu$ are plotted in Fig.~(\ref{D2minDist}d). Increasing $\mu$ fluidizes the system. As a result, it makes the cells dynamic, which rearranges frequently via non-affine displacements, leading to a broad distribution in $D^2_{min}$ as $\mu$ increases. Lastly, we plotted the $P(x_{NA})$ versus $x_{NA}$ for a range of shear rates in Fig.~(\ref{D2minDist}e). The variation in $\alpha$ is $\sim 1.48-1.86$ when we fit the curves with a non-Gaussian form. We also obtained $P(D^2_{min})$ for each of the $\dot{\gamma}$ and found that as we decrease $\dot{\gamma}$, $P(D^2_{min})$ broadens and the peak positions shifts towards larger $D^2_{min}$ values as shown in Fig.~(\ref{D2minDist}f). This is because as $\dot{\gamma}$ decreases, the system has enough time to relax and rearrange more slowly through non-affine movements, leading to an increase in the $D^2_{min}$ values.  We show the fit for the small $D^2_{min}$ part with red lines and the large $D^2_{min}$  part with blue lines in Fig.~(\ref{D2minDist}) using Eq.~(\ref{D2minTheo}) (from the main text) in (b), (d), and (f).

	\section{Steady state shear flow}
	Following  Yamamoto and Onuki \cite{Ryoichi1998}, we defined the displacement vector as:
	 \begin{equation}
	 	\label{newdisplacement}
	 	\Delta \vec{r}_i(t) = \vec{r}_i(t) - \vec{r}_i(0) - \dot{\gamma}\int_{0}^{t} y_i(\tilde{t})d\tilde{t} \hat{x},
	 \end{equation}
	 where $\hat{x}$ is the unit vector along the flow direction and $\tilde{t}$ is an auxiliary variable. The plot of the mean square displacement of the cells, $\Delta r^2(t) =\Big \langle \sum_{i = 1}^{N} \Delta \vec{r}_i(t)^2 \Big\rangle$ in Fig.~(\ref{SSrelaxStress}a) shows that as  $\dot{\gamma}$ decreases, the $\Delta r^2(t)$ approaches the $\dot{\gamma} = 0$ data (solid red line in Fig.~(\ref{SSrelaxStress}a)). The extracted the diffusion constant, $D(\phi,\dot{\gamma})$, from the slope of the long-time linear part of the mean square displacement shows that $D(\phi,\dot{\gamma})^{-1}$ scales as $\sim\dot{\gamma}^{-0.68}$ (Fig.~(\ref{SSrelaxStress}b)). Interestingly, the shear viscosity, $\eta(\phi, \dot{\gamma}) = P_{xy}(\phi, \dot{\gamma})/\dot{\gamma}$ for the same set of $\phi$ also scales as $\sim\dot{\gamma}^{-0.68}$ for the larger $\dot{\gamma}$. 
     
     To characterize the relaxation dynamics of the sheared system, we calculated $F_s(k,t)$ using
     \begin{equation}\label{Fskt}
	 	F_s(k,t)  = \frac{1}{N}\big\langle \sum_{i=1}^N \exp[-i\vec{k}.\Delta \vec{r}_i(t)]\big\rangle,
	 \end{equation}
	 where $\vec{r}_i$s are the center of the cell $i$ and $\vec{k} = k(k_x, k_y, k_z)$ is the wave vector, and $\Delta \vec{r}_i(t)$ is given in Eq.~(\ref{newdisplacement}). In the above equation, $k = \frac{2\pi}{r_{max}}$ and $r_{max}$ is the position of the first peak of the $g(r)$. The qualitative trends do not change with the choice of $k$, such as multiples of $\frac{2\pi}{r_{max}}$. Plots of $F_s(k,t)$ for $\dot{\gamma}$ ranging from $8\times10^{-5}$ to $10^{-7} s^{-1}$ in Fig.~(\ref{SSrelaxStress}d) show that $F_s(k,t)$ decays rapidly at larger $\dot{\gamma}$. As $\dot{\gamma}$ decreases, $F_s(k,t)$ approaches the zero shear value (shown as a dotted line in Fig.~(\ref{SSrelaxStress}d)).  The relaxation time, $\tau_{\alpha}(\phi,\dot{\gamma})$, calculated using $F_s(k,\tau) = 1/e$, decays  for five  $\phi$ values,  scales as $\sim\dot{\gamma}^{-0.68}$ for the larger $\dot{\gamma}$s. Hence, $D(\phi,\dot{\gamma})^{-1}$, $\tau_{\alpha}(\phi,\dot{\gamma})$, and $\eta(\phi,\dot{\gamma})$ all scale as $\sim\dot{\gamma}^{-\kappa}$ with $\kappa = 0.68$.
	
    We computed steady-state shear, $\sigma_{SS}$, and plotted it against $\dot{\gamma}$ with $X$ in logarithmic scale in Fig.~(\ref{SSrelaxStress}f). We fit the curves using Eq.~(\ref{HBEq}) (in the main text), i.e., the Herschel–Bulkley (HB) equation \cite{Herschel1926}. The lines in Fig.~(\ref{SSrelaxStress}f) are the fits with Eq.~(\ref{HBEq}) (in the main text), and the symbols are from the simulations for various $\phi$. It is clear from the figure that the steady-state shear stress fits well with the HB form with the same exponent, $n = 0.4$, as in the transient state. We denote $\sigma_P^0(\phi) \equiv \sigma_{SS}^0(\phi)$ and $K(\phi) \equiv K^{SS}(\phi)$ as steady state values. Similarly to the transient state, we multiply the $X$ axis by $\tau_c^{SS} = \big(\frac{K^{SS}(\phi)}{\sigma_{SS}^0(\phi)}\big)^{1/0.4}$ and divide the $Y$ axis by $\sigma_{SS}^0(\phi)$, and obtain a master curve as shown in the inset of Fig.~(\ref{SSrelaxStress}f).
	
	\subsection{Distribution of viscosity, \texorpdfstring{$\eta$}{eta}} To understand the mechanical behavior of the cells, we calculated the distribution of the viscosity, $\eta(\phi, \dot{\gamma}) = P_{xy}(\phi, \dot{\gamma})/\dot{\gamma}$ (measured in Pa.s), for a set of $\phi$  (see Fig.~(\ref{distVisSS}a)). As $\phi$ increases, the peak of the distribution shifts to the right and the width broadens. The distribution of $\tilde{\eta}(\phi,\dot{\gamma}) ={\eta}(\phi,\dot{\gamma}) -\langle {\eta}(\phi,\dot{\gamma}) \rangle$ in Fig.~(\ref{distVisSS}b) shows  that as $\dot{\gamma}$ increases, the peak of the distribution becomes sharper. We know that decreasing $\phi$ or increasing $\dot{\gamma}$ makes the layer fluid-like with less variation in viscosity, hence a narrow distribution of viscosity.  The distributions are almost symmetric and are well fitted with a Gaussian distribution (lines in Fig.~(\ref{distVisSS})). This result is in line with the experimental observation by McCord and Notbohm \cite{McCord2025}, where they perturbed the actomyosin contractility of collectively migrating MDCK and HaCaT cell monolayers. It was found that the actomyosin cytoskeleton and cell–cell adhesions significantly impact viscosity. In the presence of Blebbistatin, a myosin II inhibitor, actomyosin contractility is reduced, thus lowering the effective viscosity of the tissue and making the cell layer behave more like a fluid. With Blebbistatin treatment, the distribution of effective viscosity narrowed, which is in consistent with our observation when we increase $\dot{\gamma}$ or decrease $\phi$.
    
	\section{Active Brownian particle  simulations}
    In order to assess if the results are a consequence of a lack of correlation in the active noise, we simulated Active Brownian Particles (ABP). Instead of the stochastic activity, we added a term $v_0 \vec{e}$ to Eq.~(\ref{EoM}), where $v_0$ is the self-propulsion speed (measured in $\mu m/s$). The time evolution of $\vec{e}$ is given by,
    \begin{equation}
	\dot{\vec{e}}_i =\sqrt{2 D_r} \vec{\xi_r} \times {\vec{e}}_i,
	\label{evecEvol}
\end{equation}
where, $D_r$ is the rotational diffusivity (measured in $s^{-1}$) or the inverse of persistence time ($\tau_p$), $\vec{\xi_r}$ Gaussian white-noise term. We first equilibrated the ABP system in the absence of shear to prepare the initial configurations. In the shear simulations, we investigated the effects of $v_0$ and $D_r$ on the yielding transitions.  
    
At a fixed $D_r = 0.001 s^{-1}$ and $\dot{\gamma} = 5\times10^{-5} s^{-1}$, the yield stress decreases as $v_0$ increases (Fig.~(\ref{ABPDataPlot}a)). We then obtained the stress-strain profile as a function of shear rate with $v_0 = 0.001 \mu m/s$ and $D_r = 0.001 s^{-1}$ fixed (Fig.~(\ref{ABPDataPlot}b)). The the yield stress, $\sigma_P(v_0, \dot{\gamma})$, calculated using the stress-strain profiles for other $v_0$ values the yield stress, also follows the Herschel–Bulkley form and are also well fit by Eq.~(\ref{HBEq}) with the same exponent $n = 0.4$ like $\phi$ (Fig.~(\ref{ABPDataPlot}c)). The variables in Eq.~(\ref{HBEq}) and Eq.~(\ref{HBEqCollapse}) are the same except $\phi$ is now replaced by $v_0$. In addition, the data in Fig.~(\ref{ABPDataPlot}c) can be collapsed when we plot $\sigma_P(v_0, \dot{\gamma})/\sigma_P^0(v_0)$ vs. $\tau_c \dot{\gamma}$, where $\tau_c = \big(\frac{K(v_0)}{\sigma_P^0(v_0)}\big)^{1/0.4}$ is a $v_0$-dependent timescale (Fig.~(\ref{ABPDataPlot}d)).  The yielding properties as a function of $D_r$ at a fixed $v_0 = 0.001\mu m/s$ and a fixed $\dot{\gamma} = 5\times10^{-5} s^{-1}$ has a negligible effect (Fig.~(\ref{ABPDataPlot}e).  The yield stress, $\sigma_P(D_r, \dot{\gamma})$, is plotted against $\dot{\gamma}$ in Fig.~(\ref{ABPDataPlot}f).

    \newpage
    \begin{figure}[htbp]
		\begin{minipage}{\textwidth} 
        \centering
			\includegraphics[width=\textwidth]{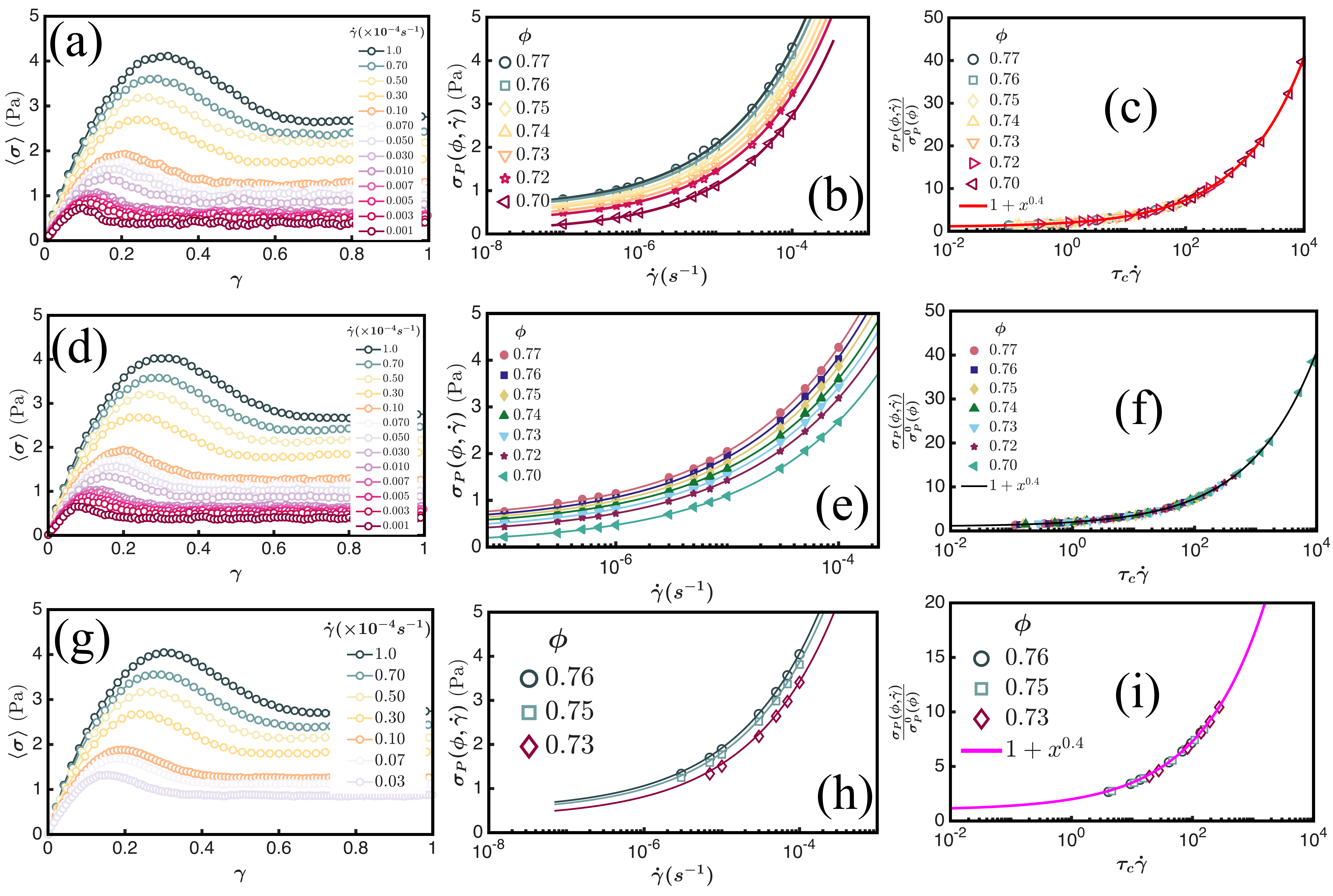}
			\caption{\textbf{Finite size effects:} (a) Ensemble-averaged stress-strain profiles at $\phi = 0.76$ and $\mu = 0.045 \mu m/\sqrt{s}$ for decreasing $\dot{\gamma}$ for $N = 500$. The system reaches a yield stress before reaching the steady state. (b) $\sigma_P(\phi,\dot{\gamma})$ versus $\dot{\gamma}$ for various $\phi$. Symbols are the simulation data, and the solid lines are power law fits to Eq.~(\ref{HBEq}) (in the main text) with $n = 0.4$. (c) Scaling collapse of the data in Fig.~(\ref{syssize}b). The solid line is the plot of the function $f(x) = 1 + x^{0.4}$. (d, e, f) Same data as (a,b,c) for $N = 1000$. (g,h, i) Same data as (a,b,c) for $N = 5000$. Other simulation parameters are: $E = 0.001$ MPa, $\mu = 0.045 \mu m/\sqrt{s}$, $\Sigma = 8.5\%$. Fig.(a,d,g) are for $\phi = 0.76$. We have reproduced Fig. (e-f) from the main text to show the comparison with other system sizes.}
			\label{syssize}
		    \end{minipage}
	\end{figure}
    \begin{figure}[htbp]
		\begin{minipage}{0.9\textwidth} 
        \centering
			\includegraphics[width=\textwidth]{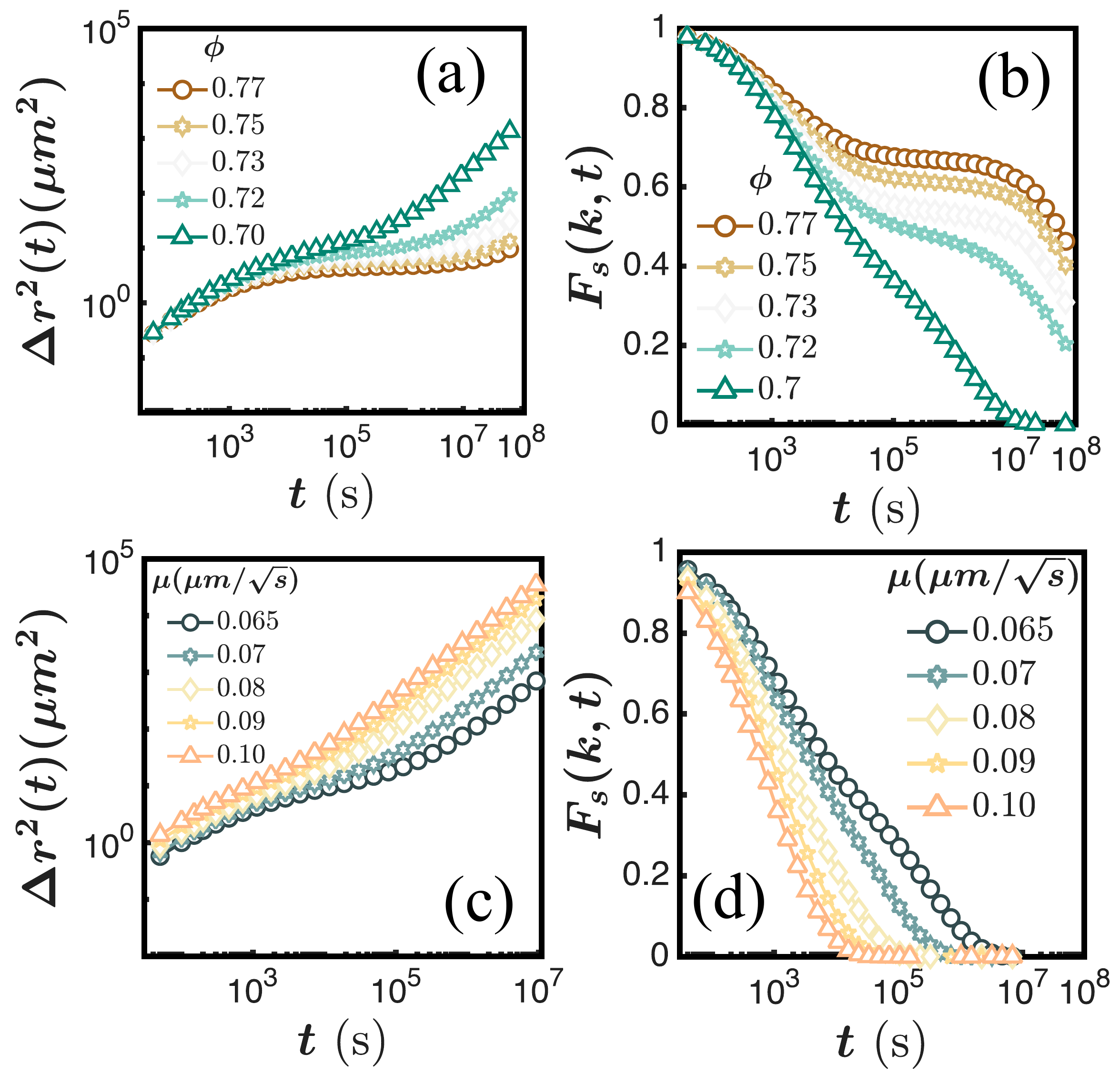}
			\caption{\textbf{Dynamics without shear}: (a) Mean square displacement, $\Delta r^2(t)$ as a function of time for five different $\phi$. As $\phi$ increases, the plateau in $\Delta r^2(t)$ becomes longer. (b) Self-intermediate scattering function, $F_s(k,t)$, as a function of time for the same set of $\phi$. In Fig.(a-b), other simulation parameters are: $E = 0.001MPa$, $\Sigma = 8.5\%$, and $\mu = 0.045 \mu m/\sqrt{s}$. (c) Mean square displacement, $\Delta r^2(t)$ as a function of time, of cells for five $\mu$ values. (d) $F_s(k,t)$ as a function of time for the same set of $\mu$. As $\mu$ increases, the decay of $F_s(k,t)$ becomes more rapid. For (c-d), other simulation parameters are: $E = 0.001$ MPa, $\Sigma = 8.5\%$, and $\phi = 0.77$.} 
			\label{woshearDynaprop}
             \end{minipage}
	\end{figure}
    
\begin{figure}[htbp]
		\begin{minipage}{0.9\textwidth} 
        \centering
			\includegraphics[width=\textwidth]{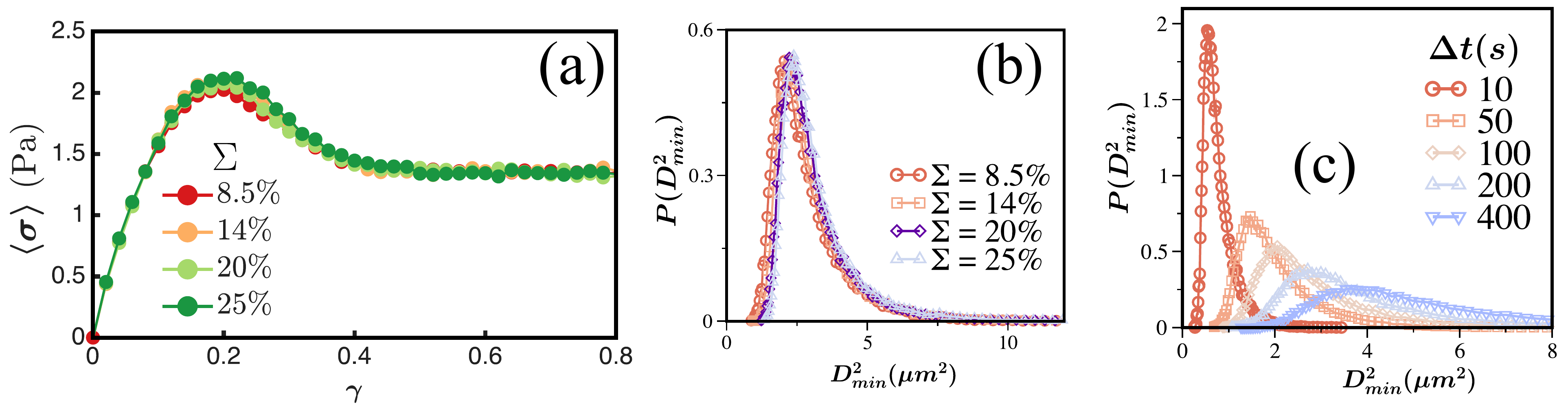}
			\caption{\textbf{Effect of polydispersity on yielding and $P(D^2_{min})$:} (a) Ensemble averaged stress-strain profiles for varying $\Sigma$. (b) $P(D^2_{\min})$ versus $D^2_{\min}$ for four  $\Sigma$ values. Changing $\Sigma$ shows little variation in the stress-strain profiles and $P(D^2_{min})$. (c) $P(D^2_{min})$ versus $D^2_{min}$ for various $\Delta t$ at $\gamma = 0.2$. The simulation parameters are: $E = 0.001$ MPa, $\mu = 0.045 \mu m/\sqrt{s}$, $\phi = 0.77$, and $\dot{\gamma} = 10^{-5} s^{-1}$.} 
			\label{Polydiseffect}
		\end{minipage}
	\end{figure}
    \begin{figure}[p]
    \centering
    \includegraphics[width=\textwidth]{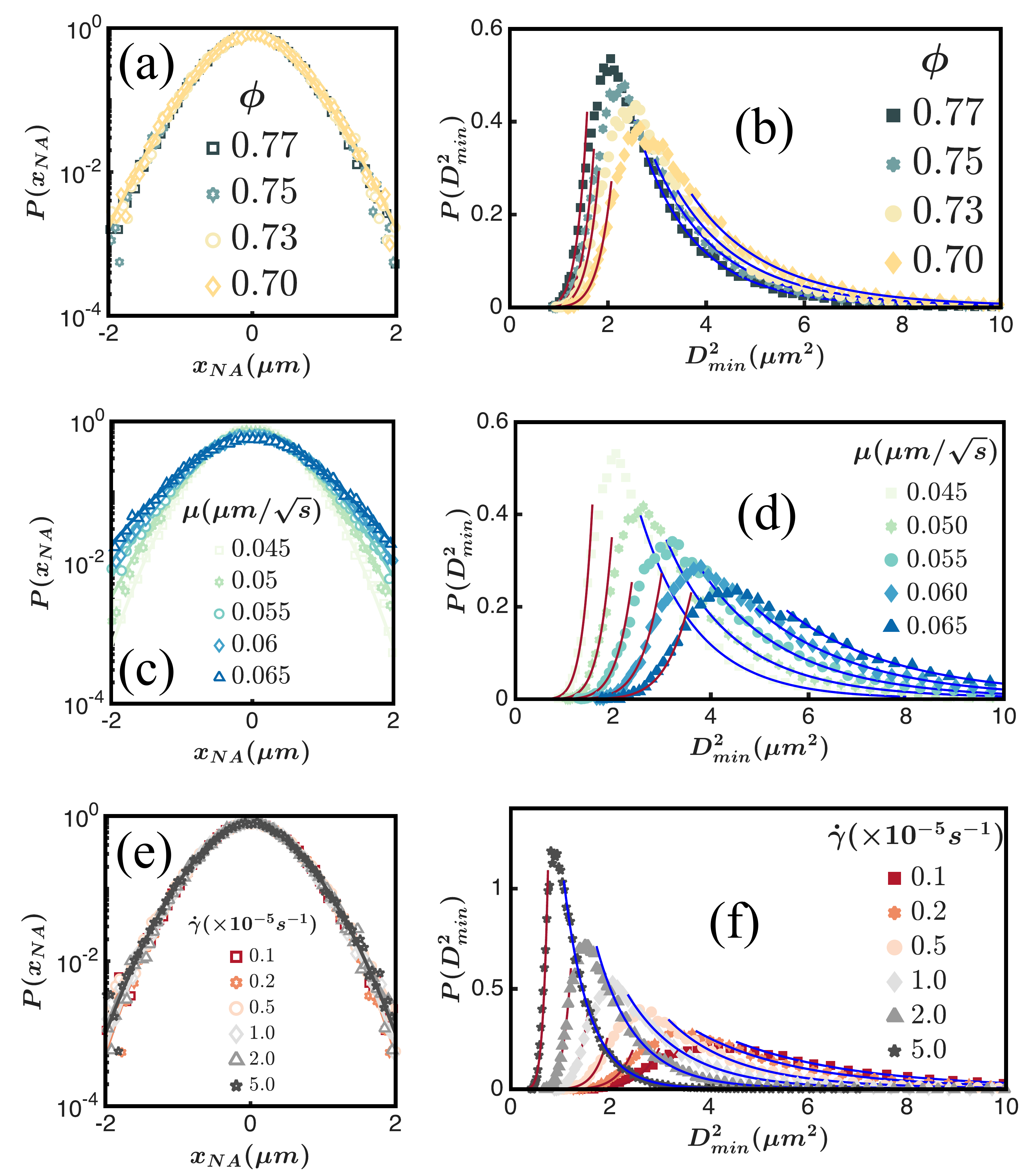}
\end{figure}
\clearpage
\begin{figure}[p]
\begin{minipage}{\textwidth} 
        \centering
    			\caption{ {\bf Distribution of the non-affine displacements as a function of parameters}: (a) $P(x_{NA})$ versus $x_{NA}$ in semi-log scale for $\phi = 0.70, 0.73, 0.75, 0.77$. The lines are the fit using Eq.~(\ref{PxNAEq}).  The variation of $\alpha$ is small, $\sim 1.53-1.56$. (b) $P(D^2_{\min})$ versus $D^2_{\min}$ obtained at various values of $\phi$. As $\phi$ decreases, the distribution broadens because more non-affine rearrangements occur in the liquid-like regime. For (a) and (b) $E = 0.001$MPa, $\dot{\gamma} = 10^{-5}s^{-1}$, $\mu = 0.045 \mu m/\sqrt{s}$, $\Sigma = 8.5\%$. (c) $P(x_{NA})$ versus $x_{NA}$ in semi-log scale for five $\mu$. Note that $\alpha$ varies between $\sim 1.86-1.39$. (d) $P(D^2_{\min})$ versus $D^2_{\min}$ obtained at various values of $\mu$. As $\mu$ increases, the peak height decreases. In (c) and (d) $\phi = 0.77$, $\dot{\gamma} = 10^{-5} s^{-1}$, $E = 0.001$MPa, $\Sigma = 8.5\%$. We choose the waiting time to be $\Delta t = 100s$. $D^2_{min}$ is measured instantaneously at a spacing of $\Delta t$. (e) $P(x_{NA})$ versus $x_{NA}$ in semi-log scale for various $\dot{\gamma}$. There is a variation in $\alpha \sim 1.48-1.86$. (f) $P(D^2_{\min})$ versus $D^2_{\min}$ at $\gamma = 0.2$ for  different shear rates. A smaller rate exhibits decreased peak height with a long-tailed distribution, whereas a larger rate produces a sharp peak. For (e) and (f) $\phi = 0.77$, $E = 0.001$MPa, $\Sigma = 8.5\%$, $\mu = 0.045 \mu m/\sqrt{s}$, $\Delta t = 500s$. $\dot{\gamma}$ is fixed at $10^{-5}s^{-1}$ for all the plots except (e) and (f).   The lines in (a), (c),  and (e) are the fit with the $P(x_{NA}) = A_1 \exp(-B_1|x_{NA}|^\alpha)$ form, where $A_1, B_1$, and $\alpha$ are the fitting parameters. We have shown the fits for the small $D^2_{min}$  part with red lines and the large $D^2_{min}$  part with blue lines using Eq.~(\ref{D2minTheo}) in (b), (d), and (f).} 
                \label{D2minDist}
                \end{minipage}
\end{figure}
\clearpage
\begin{figure}[p]
    \centering
    \includegraphics[width=\textwidth]{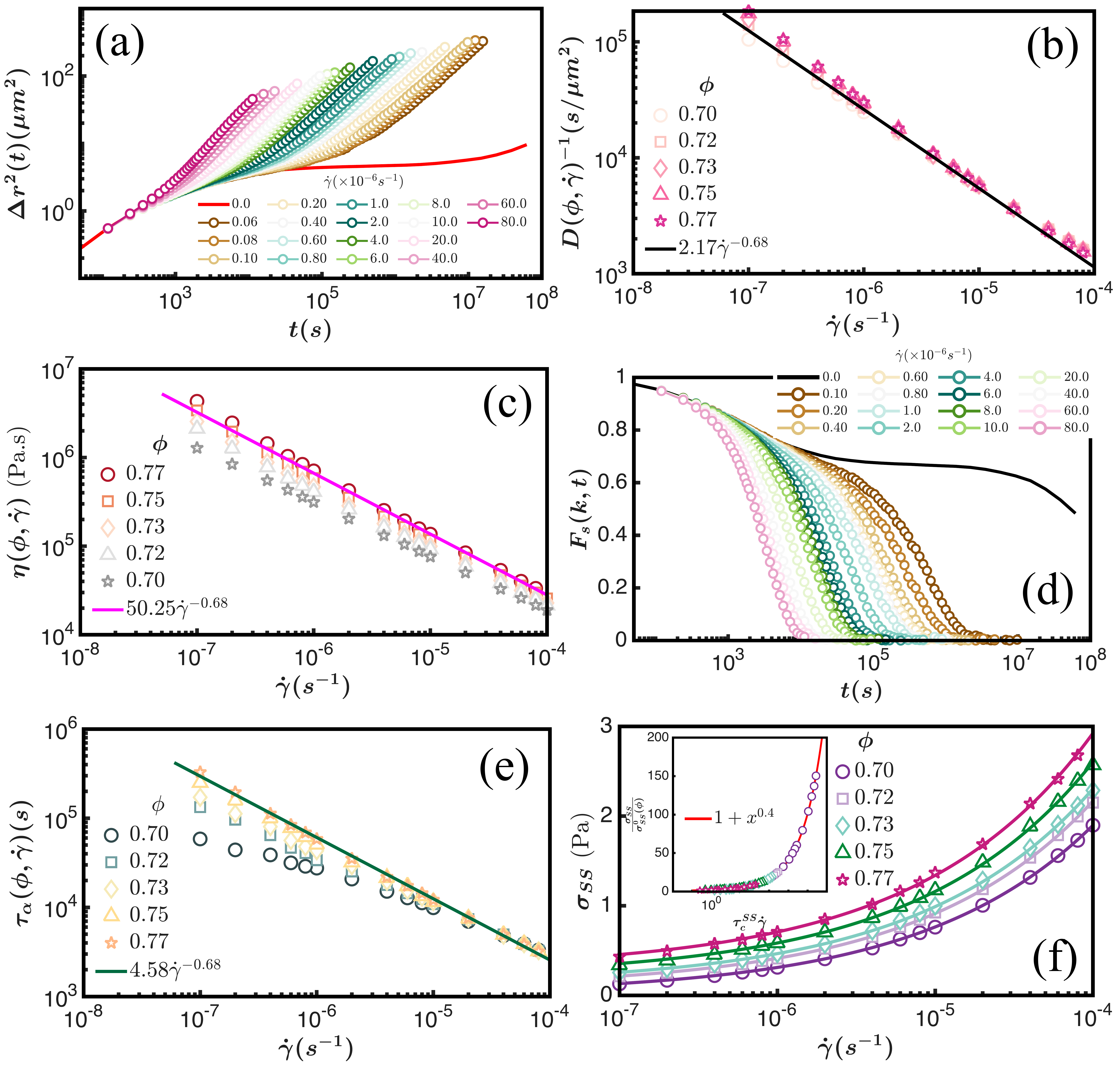}
\end{figure}
\clearpage
\begin{figure}[htbp]
		\begin{minipage}{0.9\textwidth} 
        \centering
			\caption{ {\bf Steady State Properties}: (a) Mean square displacement, $\Delta r^2(t)$ of the sheared system for $\phi = 0.77$ for a set of $\dot{\gamma}$. As $\dot{\gamma}$ decreases, the plateau of  $\Delta r^2(t)$ increases and approaching the $\dot{\gamma} = 0$ curve (red solid line). (b) Inverse diffusion constant, $D(\phi, \dot{\gamma})^{-1}$, extracted from the long time linear part of $\Delta r^2(t)$, as a function of $\dot{\gamma}$ for five $\phi$ values follows a power law scaling with $D(\phi, \dot{\gamma})^{-1} \sim \dot{\gamma}^{-0.68}$ (c) $\eta(\phi, \dot{\gamma})$ versus $\dot{\gamma}$ for five $\phi$ values also shows a power law scaling $\eta(\phi, \dot{\gamma}) \sim \dot{\gamma}^{-0.68}$.  (d) $F_s(k,t)$ as a function of $t$ for a range of $\dot{\gamma}$: going from left to right, $\dot{\gamma}$ decreases, and $F_s(k,t)$ approaches to $\dot{\gamma} = 0$ curve (black dotted line). (e) $\tau_{\alpha}(\phi, \dot{\gamma})$ versus $\dot{\gamma}$ as a function of $\phi$. $\tau_{\alpha}(\phi, \dot{\gamma})$ saturates as $\dot{\gamma}$ decreases. $\tau_{\alpha}(\phi, \dot{\gamma}) \sim \dot{\gamma}^{-0.68}$ for large $\dot{\gamma}$. (f) Steady state stress, $\sigma_{SS}$, versus $\dot{\gamma}$ in log-scale for the same set of $\phi$ values. The lines are fit using Eq.~(\ref{HBEq}) with the same exponent, $0.4$, as the transient state. Inset: The master curve after proper scaling of the main figure (f) (see text).} 
			\label{SSrelaxStress}
		\end{minipage}
	\end{figure}
	\begin{figure}[htbp]
		\begin{minipage}{\textwidth} 
        \centering
			\includegraphics[width=1\columnwidth]{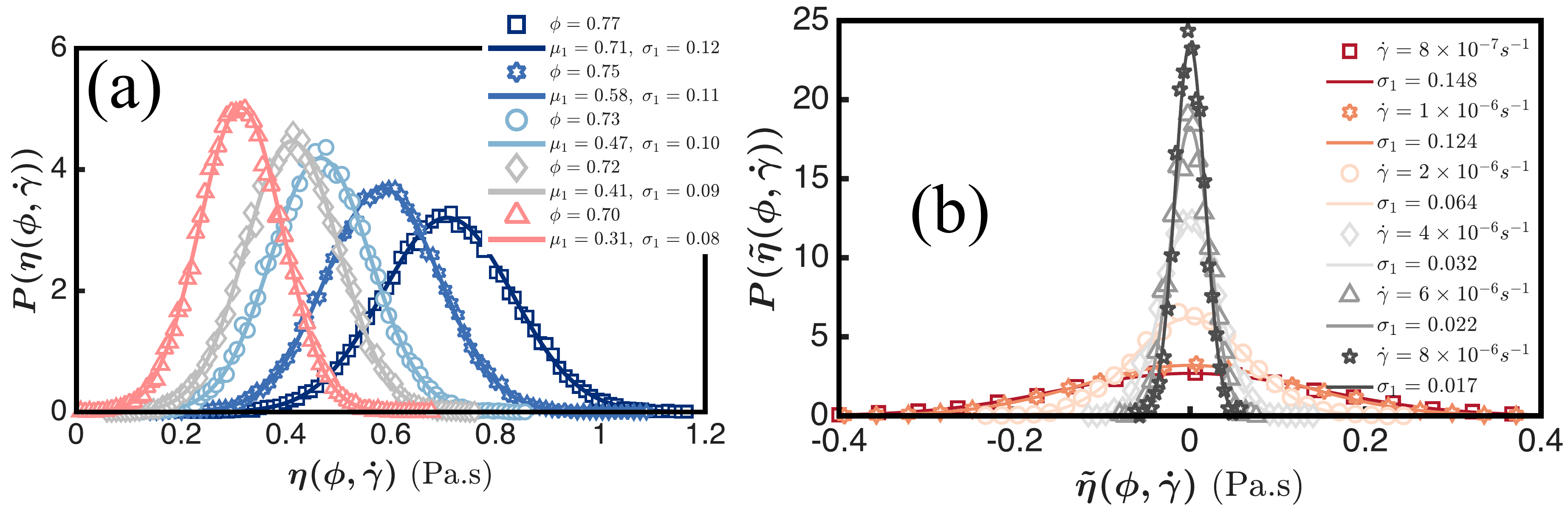}
			\caption{ {\bf Distribution of steady state viscosity}: (a) Distribution of ${\eta}(\phi,\dot{\gamma})$ for various $\phi$. As $\phi$ decreases, the peak of $P({\eta}(\phi,\dot{\gamma}))$ increases and the width of the distribution ($\sigma_1$) decreases. Symbols are the simulation data, and the lines are the fits with a Gaussian distribution with mean $\mu_1$ and standard deviation $\sigma_1$. (b) Distribution of $\tilde{\eta}(\phi,\dot{\gamma}) = {\eta}(\phi,\dot{\gamma}) - \langle {\eta}(\phi,\dot{\gamma}) \rangle$ for a range of $\dot{\gamma}$ for $\phi = 0.77$. Decreasing $\dot{\gamma}$, broadens the distribution. Symbols are the simulation data, and the lines are the fits with a Gaussian distribution standard deviation $\sigma_1$.} 
			\label{distVisSS}
		\end{minipage}
	\end{figure}

\begin{figure}[htbp]
		\begin{minipage}{\textwidth} 
        \centering
			\includegraphics[width=1\columnwidth]{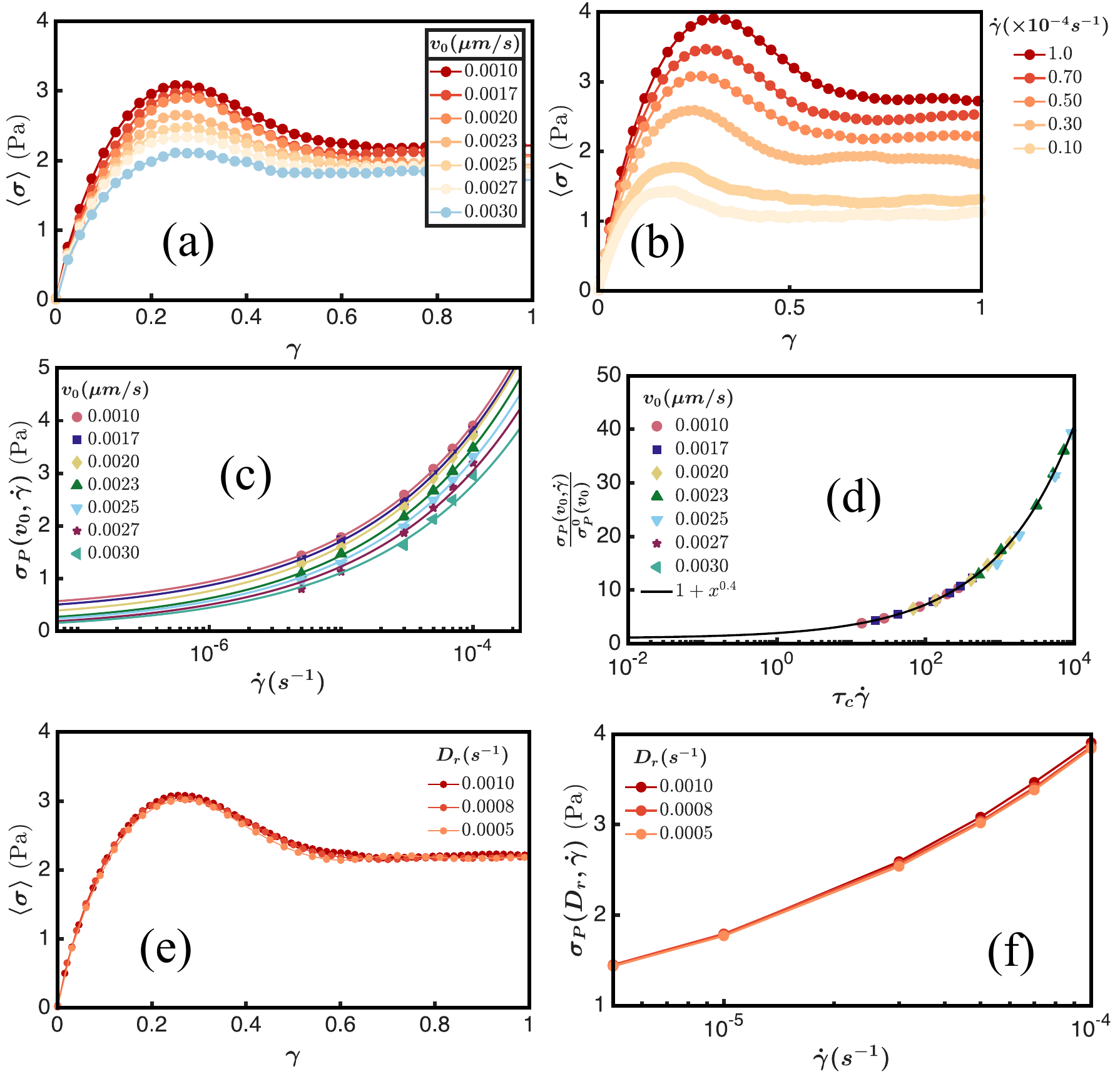}
            \caption{ {\bf Yielding in the ABP model}: (a) Ensemble-averaged stress-strain profiles for increasing value of $v_0$ from top to bottom at a fixed $D_r = 0.001 s^{-1}$ and $\dot{\gamma} = 5\times10^{-5} s^{-1}$. (b) Ensemble-averaged stress-strain profiles for decreasing $\dot{\gamma}$ at a fixed $v_0 = 0.001\mu m/s$ and  $D_r = 0.001 s^{-1}$. (c) Yield stress, $\sigma_{P}(v_0, \dot{\gamma})$ versus $\dot{\gamma}$. Symbols are the simulation data and solid lines are power law fits to Eq.~(\ref{HBEq}) with $n = 0.4$. (d) Collapse of the data in Fig.~(\ref{ABPDataPlot}c). The solid black line is the plot of the function $f(x) = 1 + x^{0.4}$ (Eq.~(\ref{HBEqCollapse})). (e) Ensemble-averaged stress-strain profiles as a function of $D_r$ for a fixed $v_0 = 0.001 \mu m/s$  $\dot{\gamma} = 5\times 10^{-5} s^{-1}$. (f) $\sigma_P(D_r,\dot{\gamma})$ versus $\dot{\gamma}$ for three $D_r$ values. Other parameters in the simulations are: $\phi = 0.75$, $\Sigma = 8.5\%$, $E = 0.001$MPa, $\mu = 0$.} 
			\label{ABPDataPlot}
		\end{minipage}
	\end{figure}

\end{document}